\documentclass[amsmath,amssymb,reprint,aps,floatfix,prx]{revtex4-1}

\usepackage{graphicx}
\usepackage{amssymb}
\usepackage{amsmath}
\usepackage{epstopdf}
\usepackage{xcolor}
\usepackage{url}

\usepackage{booktabs}
\usepackage{array}
 \newcolumntype{M}{>{$\displaystyle}l<{$}}
  \newcolumntype{N}{>{$\displaystyle}c<{$}}
  \newcolumntype{O}[1]{>{\arraybackslash}m{#1}}

\usepackage{colortbl}
\makeatletter

    \def\CT@@do@color{%
      \global\let\CT@do@color\relax
            \@tempdima\wd\z@
            \advance\@tempdima\@tempdimb
            \advance\@tempdima\@tempdimc
    \advance\@tempdimb\tabcolsep
    \advance\@tempdimc\tabcolsep
    \advance\@tempdima2\tabcolsep
            \kern-\@tempdimb
            \leaders\vrule
                    \hskip\@tempdima\@plus  1fill
            \kern-\@tempdimc
            \hskip-\wd\z@ \@plus -1fill }
    \makeatother

\definecolor{bg}{RGB}{230,230,230}

\usepackage[utf8]{inputenc}
\usepackage{cleveref}
\crefname{section}{§}{§§}
\Crefname{section}{§}{§§} 
 
\usepackage{colortbl}
\definecolor{bg}{RGB}{230,230,230}

\newcommand{\be}{\begin{equation}}
\newcommand{\ee}{\end{equation}}
\newcommand{\bi}{\begin{itemize}}
\newcommand{\ei}{\end{itemize}}
\newcommand{\bdm}{\begin{displaymath}}
\newcommand{\edm}{\end{displaymath}}
\newcommand{\bea}{\begin{eqnarray}}
\newcommand{\eea}{\end{eqnarray}}
\newcommand{\bml}{\begin{multline}}
\newcommand{\emult}{\end{multline}}
\newcommand{\nn}{\nonumber}

\newcommand{\bs}{\boldsymbol}
\newcommand{\bsu}{\boldsymbol{u}}
\newcommand{\bsO}{\boldsymbol{\Omega}}
\newcommand{\bso}{\boldsymbol{\omega}}
\newcommand{\bsg}{\boldsymbol{g}}

\newcommand{\Pef}{Pe_{ \!  f \! \!  f}}

\newcommand{\PeTW}{Pe_{TW}}
\newcommand{\ReTW}{Re_{TW}}
\newcommand{\Ref}{Re_{ \! f \! \!  f}}

\newcommand{\RoC}{Ro_{\mathcal{C}}}
\newcommand{\uf}{U_{ \! f \! \!  f}}

\newcommand{\uTW}{U_{TW}}
\newcommand{\tauTW}{\tau_{\scriptscriptstyle TW}}
\newcommand{\tauf}{\tau_{ \! f \! \!  f}}
\newcommand{\wRa}{\widetilde{Ra}}
\newcommand{\Rol}{Ro_\ell}

\begin{document}

\title{Connections between Non-Rotating, Slowly Rotating, and Rapidly Rotating Turbulent Convection Transport Scalings} 

\author{Jonathan M. Aurnou} 
\email[]{aurnou@g.ucla.edu}
\affiliation{University of California at Los Angeles, Los Angeles, California 90095-1567 USA}
\author{Susanne Horn}
\affiliation{Coventry University, Coventry CV1 5FB, United Kingdom}
\author{Keith Julien}
\affiliation{University of Colorado at Boulder, Boulder, Colorado 80309, USA}

\date{\today}

\begin{abstract}
In this study, we investigate and develop scaling laws  as a function of external non-dimensional control parameters for heat and momentum transport for non-rotating, slowly rotating and rapidly rotating turbulent convection systems, with the end goal of forging connections and bridging the various gaps between these regimes. Two perspectives 
are considered, one where turbulent convection is viewed from the standpoint of an applied temperature drop across the domain and the other with a viewpoint in terms of an applied heat flux. While a straightforward transformation exist between the two perspectives indicating equivalence, it is found the former provides a clear set of connections that bridge between the three regimes. Our generic convection scalings, based upon an Inertial-Archimedean balance, produce the classic diffusion-free scalings for the non-rotating limit (NRL) and the slowly rotating limit (SRL). This is characterized by a free-falling fluid parcel on the global scale possessing a thermal anomaly on par with the temperature drop across the domain.  In the rapidly rotating limit (RRL), the generic convection scalings are based on a Coriolis-Inertial-Archimedean (CIA) balance, along with a local fluctuating-mean advective temperature balance. This produces a scenario in which anistropic fluid parcels attain a thermal wind velocity and where the thermal anomalies are greatly attenuated compared to the total temperature drop. 

We find that turbulent scalings may be deduced simply by consideration of the generic non-dimensional transport parameters --- local Reynolds $Re_\ell = U \ell /\nu$; local P\'eclet $Pe_\ell =  U \ell /\kappa$; and Nusselt number $Nu = U \vartheta/(\kappa \Delta T/H)$ --- through the selection of physically relevant estimates for length $\ell$,  velocity $U$ and temperature scales $\vartheta$ in each regime. Emergent from the scaling analyses is a unified continuum based on a single external control parameter, the convective Rossby number $\RoC = \sqrt{g \alpha \Delta T / 4 \Omega^2 H}$, that strikingly appears in each regime by consideration of the local, convection-scale Rossby number $\Rol=U/(2\Omega \ell)$. Thus we show that $\RoC$ scales with the local Rossby number $\Rol$ in both the slowly rotating and the rapidly rotating regimes, explaining the ubiquity of $\RoC$ in rotating convection studies. We show in non-, slowly,  and rapidly rotating systems that the convective heat transport, parametrized via $Pe_\ell$, scales with the total heat transport parameterized via the Nusselt number $Nu$. Within the rapidly-rotating limit, momentum transport arguments generate a scaling for the system-scale Rossby number, $Ro_H$, that, recast in terms of the total heat flux through the system, is shown to be synonymous with the classical flux-based `CIA' scaling, $Ro_{CIA}$.  These, in turn, are then shown to asymptote to $Ro_H \sim Ro_{CIA} \sim \RoC^2$, demonstrating that these momentum transport scalings are identical in the limit of rapidly rotating turbulent heat transfer.
\end{abstract}

\maketitle

\subsection*{Popular Summary}
Buoyancy-driven convection is likely the dominant driver of turbulent motions in the universe, and thus, is widely studied by physicists, engineers, geophysicists and astrophysicists.  Maybe unsurprisingly, these different communities discuss the gross convective behaviors in different ways, often without significant cross-talk existing between them.  Here, we seek to draw connections between these communities. We do so by carrying out a set of basic scale estimations for how heat and fluid momentum transport should behave in non-rotating, slowly rotating and rapidly rotating buoyancy-driven convective environments.  We find that slowly and rapidly rotating scalings can be inter-related via one parameter, the so-called convective Rossby number $\RoC$, a dissipation-free parameter measuring the importance of buoyancy driving relative to rotation. Further, we map between non-flux-based and the flux-based,  buoyancy-driven scalings used by different groups.  In doing so, these scalings show that there are clean connections between the different communities' approaches and that a number of the seemingly different scalings are actually synonymous with one another.

\section{Introduction}
Accurate parameterizations are ubiquituously sought for the turbulent transport properties of fluid dynamical systems. In buoyancy-driven convection systems, the heat and momentum transport properties are the main focii of such investigations \citep{Ahlers09, KingBuffett13, Grooms14, Plumley19}.  These transport estimates are essential for understanding the possible behaviors of a given system, and for extrapolating these behaviors to extreme industrial, geophysical and astrophysical settings that are difficult to simulate directly \citep[e.g.,][]{Stevenson79, GAG02, Diamond09, RobertsKing13, Gurcan15, Featherstone16spectral, Cao18, Wood13, Horn18}. 

In the Rayleigh-B\'enard convection systems considered here, warmer fluid is maintained at the base of the fluid layer and colder fluid is maintained at the top of the layer, defined with respect to the gravity vector $\boldsymbol{g}$ that is parallel to the background temperature gradient. In addition, our system is rotating at angular velocity $\boldsymbol{\Omega}$ that is oriented in the axial $\boldsymbol{\hat{e}_z}$-direction. This system is gravitationally unstable, and drives buoyant convective flows across the fluid layer that advect both heat and momentum. We describe this system generally throughout this paper, but it can be thought of as an extended plane layer \citep{Julien98}, a finite cylinder \citep{Cheng18}, or a spherical shell of fluid \citep{Marti16}. 

A scaling analysis is presented using generic scales  for the characteristic fluid properties occurring in the non-rotating, slowly rotating and rapidly rotating turbulent limits. This analysis generates a large-scale, free-fall flow regime in the non-rotating and slowly rotating limits, and a small-scale, thermal wind flow in the rapidly rotating limit. The generic nature of our scaling analysis allows us to provide connections between the different regimes.  For instance, we show that the convective Rossby number, $\RoC$, arises ubiquitously in scaling estimates for turbulent rotating convection, both in the rapidly rotating and slowly rotating end-member limits.  Further, $\RoC$ is shown to be equivalent to $\Rol$, which describes the Rossby number for the rotating flow dynamics on the local convective scale.  The rotating scalings developed show how numerous heat and momemtum transport laws can all be inter-related via integer powers of $\RoC$ (or, synonymously, $\Rol$), thus providing novel ties between the different scaling regimes.

\section{Governing Equations and Parameters}
The governing equations of rotating thermal convection in a Boussinesq fluid are 
\begin{subequations}
\begin{eqnarray}
\partial_t \boldsymbol{u} + \boldsymbol{u} \cdot \nabla \boldsymbol{u}  + 2 \boldsymbol{\Omega} \times \boldsymbol{u} &=& - \nabla p + \boldsymbol{g} \alpha  \theta + \nu \nabla^2 \boldsymbol{u} , \\ 
\partial_t \theta + (\boldsymbol{u} \cdot \nabla \theta)^\prime  
&=&  (\boldsymbol{\hat{e}_g} \cdot \boldsymbol{u}) \partial_g \overline{T} + \kappa \nabla^2 \theta , \\ 
\partial_t \overline{T} + \nabla \cdot \overline{(\boldsymbol{u}  \theta)}
&=&  \kappa \nabla^2 \overline{T} , \\ 
 \nabla \cdot \boldsymbol{u} &=& 0
\end{eqnarray}
\label{GE}
\end{subequations} 
\citep[e.g.,][]{Sprague06, Calkins15rapids}. Other effects are not considered here, such as those due to magnetic fields \citep{King15}, centrifugal buoyancy \citep{Horn19}, and non-Oberbeck-Boussinesq-ness \citep{Horn14}. In the Navier-Stokes equation (\ref{GE}a), the velocity vector is $\boldsymbol{u}$; the angular rotation velocity along the axial coordinate $z$ is $\boldsymbol{\Omega}$; $p$ is the modified pressure; $\alpha$ is the thermal expansivity; $\boldsymbol{g}$ is the gravity vector; and $\nu$ is the fluid's kinematic viscocity.

Temperature is $T = \overline{T}+ \theta$ where the overbar denotes averaging over surfaces perpendicular to $\boldsymbol{g}$. Thus, $\overline{T}$ is the `laterally averaged' temperature and $\theta$ is the temperature fluctuation. Equation (\ref{GE}b) is the fluctuating temperature evolution equation and  (\ref{GE}c) describes the evolution of the laterally averaged temperature field. In the fluctuating temperature equation (\ref{GE}b),  we use the abbreviated notation $(\boldsymbol{u} \cdot \nabla \theta)^\prime = \boldsymbol{u} \cdot \nabla \theta - \nabla \cdot \overline{(\boldsymbol{u}  \theta)}$.  Convective motions in this system are driven by an unstable, system scale temperature gradient $\partial_g \overline{T} = \mathcal{O}(\Delta T / H)$ measured in the direction of gravity $\boldsymbol{\hat{e}_g}$, where $\Delta T$ is the temperature drop across the fluid layer of system depth $H$.  Here $\Delta T$ is sustained either via fixed temperature boundaries or via an applied heat flux $Q$ \citep[cf.][]{Calkins15Rapid}. Depending on the set-up, $\boldsymbol{\hat{e}_g}$ can be oriented in the axial direction $\boldsymbol{\hat{e}_z}$ \citep[e.g.,][]{Cheng15}, the cylindrically radial direction $\boldsymbol{\hat{e}_s}$ \citep[e.g.,][]{Gillet06}, or the spherically radial direction $\boldsymbol{\hat{e}_r}$ \citep[e.g.,][]{Marti16}. 

Here we take the characteristic convective velocity to be $U$, the characteristic length scale to be $\ell$, and the characteristic temperature anomaly to be $\vartheta$.  The slowly rotating limit (SRL) is defined such that the inertial forces greatly exceed the Coriolis force:
\be
\bs{u} \cdot \nabla \bs{u} \gg 2 \bs{\Omega} \times \bs{u} \quad \longrightarrow \quad \frac{U^2}{\ell} \gg 2 \, \Omega U \, .
\ee
The ratio of these terms, the so-called local Rossby number defined with the characteristic scales of the convection, is
\be
\Rol \equiv \frac{U}{2 \Omega \ell } \gg 1 \, .
\ee
In the rapidly rotating limit (RRL) of Rayleigh-B\'enard convection, the Coriolis forces dominate over the inertial forces, 
\be
\bs{u} \cdot \nabla \bs{u} \ll 2 \bs{\Omega} \times \bs{u} \quad \longrightarrow \quad \frac{U^2}{\ell} \ll 2 \, \Omega U
\ee
Thus, 
\be
\Rol \equiv \frac{U}{2 \Omega \ell } \ll 1 \, .
\ee
We note then that the local Rossby number estimates the strength of inertial advection using the estimated convective velocity and length scales considered, normalized by the Coriolis acceleration. 

%

We are interested in ascertaining turbulent scaling laws for the heat transported across the system scale $H$ and for the local momentum and heat transport carried by the fluid motions occurring on the convective scale $\ell$.  The system-scale heat transport is measured by the Nusselt number
\be
Nu =  \frac{Q H}{\rho c_P \kappa \Delta T} \, \sim \frac{U \vartheta H}{\kappa \Delta T} \, , 
\ee
where $\rho$ is the fluid's density and $c_P$ its specific heat capacity. Here $Q \sim \rho c_P U \vartheta$ is the total (superadiabatic) heat flux per unit area, which we assume is dominated by the turbulent convective transport component (i.e., $Nu \gg 1$).   The momentum and heat transported on the characteristic convective scale is estimated via the local Reynolds and P\'eclet numbers
\be
Re_\ell = \frac{U \ell}{\nu} \, , \quad Pe_\ell = \frac{U \ell}{\kappa} \, .
\ee


The $Nu$, $Re_\ell$ and $Pe_\ell$ transport scalings will be formulated in terms of equations (\ref{GE})'s non-dimensional control parameters, which are the Prandtl, Rayleigh and Ekman numbers \citep[e.g.,][]{Plumley19}.  The Prandtl number describes the fluid's thermophysical properties,
\be
Pr 
= \frac{\nu}{\kappa} \, , 
\ee 
where $\kappa$ and $\nu$ are the thermal diffusivity and kinematic viscosity, respectively. 
The Ekman number, $E$ gives the estimated ratio of system-scale viscous and Coriolis forces:
\be
E = \frac{\nu}{2 \Omega H^2} \, .
\ee 
The Rayleigh number estimates the strength of the buoyancy forcing 
\be
Ra = \frac{g \alpha  \Delta T H^3}{\nu \kappa} \, .
\ee
The non-dimensional bouyancy forcing will also be presented in three alternative forms. The first of these is in terms of the flux Rayleigh number, based on the heat flux through the system,
\be
Ra_F = Ra \, Nu = \frac{g \alpha  Q H^4}{\rho c_P \nu \kappa^2}  \, . 
\ee
Following \citet{Christensen02} and \citet{Christensen06}, the second form is given in terms of the rotating, diffusivity-free, so-called `modified flux Rayleigh number,' 
\be
Ra_F^* = \frac{Ra_F E^3}{Pr^2} = \frac{g \alpha Q}{8 \rho c_P \Omega^3 H^2} \, . 
\ee
(Although this convention is not used here, the oceanic and atmospheric communities write these flux-based parameters in terms of the buoyancy flux $\mathcal{B} = g \alpha Q / (\rho c_P)$ and the Coriolis parameter $f = 2 \Omega$ \citep[e.g.,][]{Maxworthy94}. Then $Ra_F = \mathcal{B} H^4 / \nu \kappa^2$ and $Ra_F^* = \mathcal{B} / f^3 H^2$.)
The third form is the convective supercriticality, 
\be
\widetilde{Ra} = Ra / Ra_{crit} \, ,
\ee
where $Ra_{crit}$ is the critical Rayleigh number above which buoyancy-driven fluid motions first onset in a given convection system \citep{Chandra61, Julien99, Zhang17}. Thermal convection is active whenever  $\widetilde{Ra} \geq 1$. No convection occurs for $\widetilde{Ra} < 1$, unless a subcritical branch also exists giving rise to a hysteretic bistable state. This has have been found in low $Pr$, rapidly rotating convection studies in spheres, such as in \citet{Guervilly16, Kaplan17}. The critical Rayleigh number is approximately $10^3$ in non-rotating systems \citep[e.g.,][]{Sparrow64}.  More specifically, $Ra_{crit} = 1708$ for no-slip mechanical boundary conditions in a non-rotating, horizontally-infinite layer of fluid. In contrast, in a rotating plane layer of $Pr \gtrsim 0.67$ fluid, the critical Rayleigh number is a strong function of the rotation rate and fluid viscosity: 
\be
Ra_{crit} \simeq 8.7 \, E^{-4/3} \, , 
\ee
and convection onsets in the form of stationary modes.  In lower Prandtl number fluids such that $Pr \lesssim 0.67$, convection first develops via oscillatory modes \citep{Chandra61, Horn17, Aurnou2018} and the critical Rayleigh number in a plane layer is $Ra_{crit} \simeq 17.4 \, (E/Pr)^{-4/3}$ \citep{ChandraElbert55, Julien97}. 
Thus, in plane layer geometries, $Ra_{crit}$ depends on the rotation rate and the fluid's thermal diffusivity in low $Pr$ fluids.  
In rotating spherical geometries, the onset is always to $Pr$-dependent oscillatory convection \citep{Jones00}.
Although $Pr$ can affect $Ra_{crit}$ in rotating fluids \citep[e.g.,][]{Dawes01}, it does not affect the outcome of our analyses since all the diffusion coefficients drop out of the final expressions. For simplicity, then, we will choose to consider only the moderate $Pr$ relationship $Ra_{crit} \sim E^{-4/3}$ from here onwards. 

Lastly, we present the convective Rossby number, $\RoC$, which arises ubiquitously in studies of rotating convection.  This non-dimensional parameter estimates the ratio of buoyancy and Coriolis forces and is commonly defined as  
\be
\RoC \equiv \sqrt { \frac{g \alpha \Delta T}{4 \Omega^2 H} } = \sqrt{ \frac{Ra E^2}{Pr} } \, .
\label{RoCdefn}
\ee
The convective Rossby number is taken to be the essential control parameter in many studies of rotating convection \citep[e.g.,][]{Gilman77, Julien96, Kunnen08, Weiss11, Stevens13, JQZ17, Horn18}, and is also claimed to control numerous transitions in rotating convection behavior \citep[e.g.,][]{Aurnou07, Gastine13, gastine2013solar, KingAurnou13, Soderlund14, Horn15, Mabuchi15}. Further, in many rotating convection and dynamo studies, the buoyancy forcing is parameterized in terms of the square of the convective Rossby number, although it is referred to there as the `modified Rayleigh number', $Ra^* = \RoC^2$ \citep[e.g,][]{Christensen02}. 

\begin{figure*}[t!]
\begin{center}
\centerline{\includegraphics[width=1\textwidth]{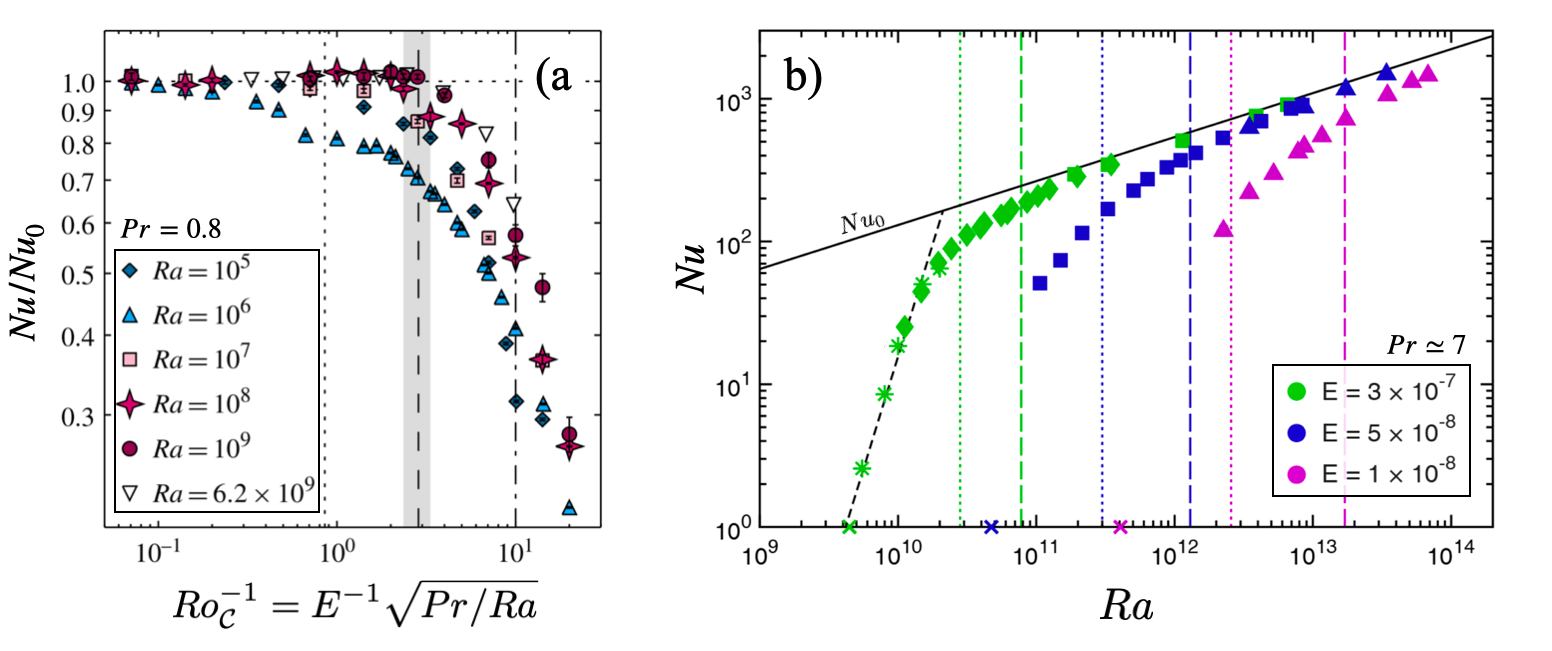}}
\caption{Rotating convection heat transport surveys carried out in the slow rotating and rapidly rotating frameworks. a) Slowly rotating approach: Fixed $Ra$ data shingles from the numerical study of \citet{Horn15}. For each data shingle, the vertical temperature difference $\Delta T \, (\propto Ra)$ is fixed and the angular rotation rate $\Omega \, (\propto E^{-1})$ is varied. b) Rapidly rotating approach: Fixed $E$ data shingles from the laboratory-numerical study of \citet{Cheng20}. Here $\Omega$ is fixed and $\Delta T$ is varied along each data shingle. The colored x-symbols mark each $E$-shingle's $Ra_{crit}$ value. The corresponding non-rotating heat transfer efficiency is denoted by $Nu_0$.}  
\label{F:ExptComp}
\end{center}
\end{figure*}
\subsubsection*{Parameter Surveys}
Within the fluid physics community, rotating convection studies often take the non-rotating limit (NRL) as their philosophical starting point. This assumes an inertial velocity scale and then the inertial turbulence is perturbed with increasing rotational effects.  Within this buoyancy-dominated framework, surveys are carried out at various fixed values of the buoyancy forcing, e.g., fixed $Ra \propto \Delta T$, while the angular rotation rate of the system $\Omega$ is systematically increased \cite[e.g.,][]{Zhong09, Weiss11, JQZ17}. An  example of this approach is shown in Figure \ref{F:ExptComp}a, which is adapted from the numerical investigation of  \citet{Horn15}.  Six different cuts through parameter space are shown, with each data ``shingle'' made at a fixed $Ra$ value as shown in the legend box \citep[e.g.,][]{Cheng16}. The control parameter displayed along the abscissa is $1/\RoC$, which in this case varies only as a function of the non-dimensional rotation rate of the system $E^{-1} \propto \Omega$.  The ordinate shows the Nusselt number, $Nu$, normalized by its NRL value at each $Ra$-value, $Nu_0(Ra)$.  

In the geophysical and astrophysical fluid dynamics communities, it is typically argued that convection occurs within the rapidly rotating limit (RRL) \citep[e.g.,][]{Calkins18}. With this guiding principle in mind, the Ekman number is typically fixed at some low value whilst $Ra$ is varied along each data shingle. Figure \ref{F:ExptComp}b, which is adapted from the laboratory-based study of \citet{Cheng20}, shows this approach well.  Three different fixed Ekman number shingles are shown. Rayleigh number values are shown on the $x$-axis and the $y$-axis denotes the Nusselt number values.  (The solid black line denotes the NRL scaling $Nu_0(Ra)$.) Small x's on the abscissa denote $Ra_{Crit} = 8.7 E^{-4/3}$, the critical $Ra$ value at which stationary planar rapidly rotating convection onsets at a given $E$ value. Such a survey uses $\wRa = Ra/Ra_{crit} = 1$ as its philosophical starting point, and then perturbs the system with ever increasing values of $\wRa$. In these studies, $\RoC$ is not used as a control parameter, but is often checked {\it a posterio} to see if it can collapse the data \cite[e.g.,][]{Gastine13, Ecke14, Cheng15, Cheng20, Hawkins20}.  

The two panels of Figure \ref{F:ExptComp} are qualitative mirror images of one another.  Starting from different ends of the inertially- versus rotationally-dominated ranges, they show nearly identical data but harvested along different slices through the same parameter spaces.  Figure \ref{F:ExptComp}a assumes a high $Ra$, slowly rotating limit (SRL) dominated by buoyancy effects, whereas Figure \ref{F:ExptComp}b assumes a low $E$, rapidly rotating limit (RRL) dominated by Coriolis forces. 

The goal of this study is to develop transport scalings that we bridge the gaps between the NRL, SRL and RRL convective world views. A particularly important finding is the relative importance of the free-fall terminal velocity in the non-rotating and slowly rotating limits and of the thermal wind terminal velocity in the rapidly rotating limit, and how these velocities are related to one another via $\RoC$.

\section{The Non-Rotating and Slow Rotating Limits}
\label{sectSRL}
\begin{figure*}[t!]
\begin{center}
\centerline{\includegraphics[width=1\textwidth]{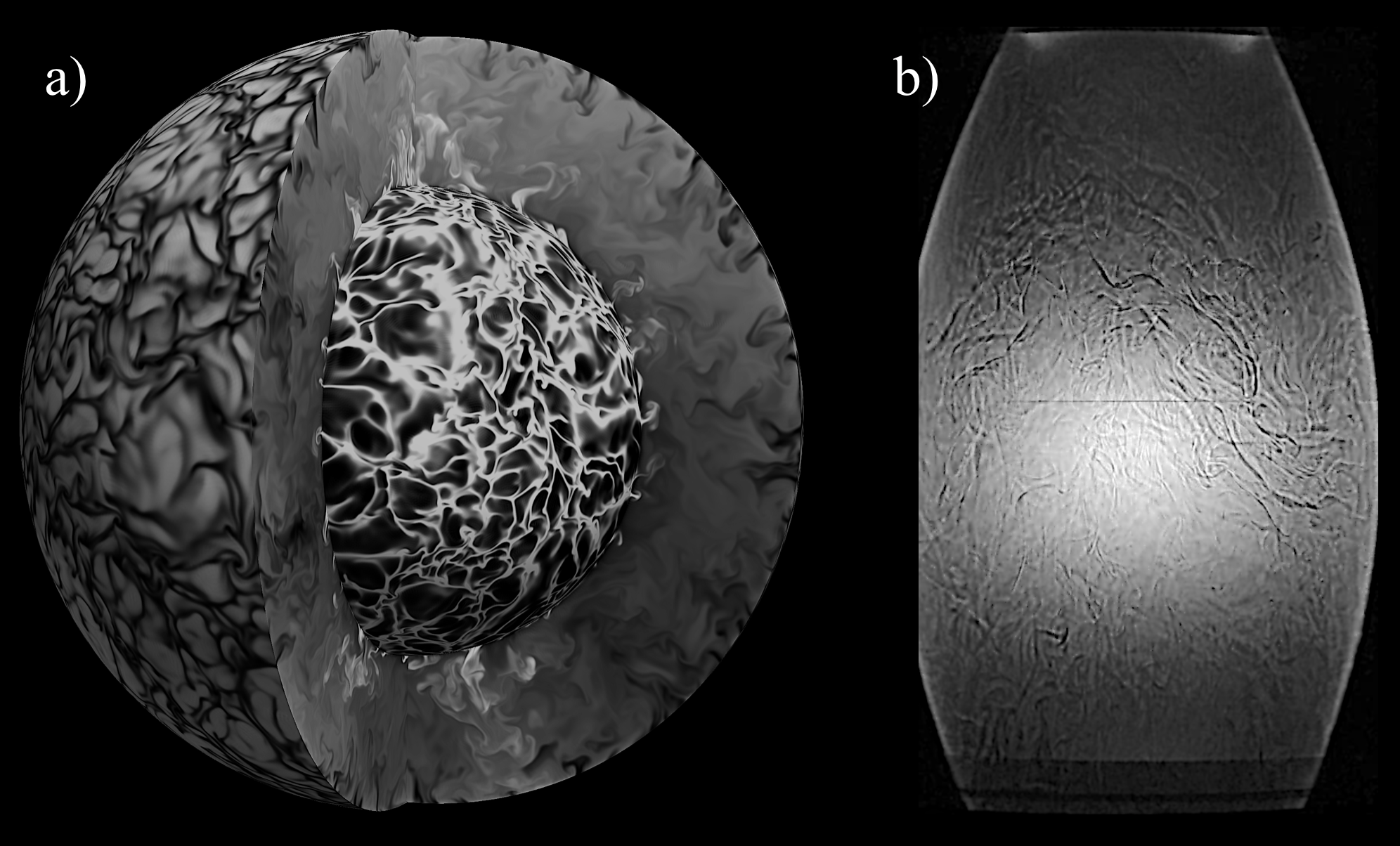}}
\caption{\small Non-rotating convective flows, which approximate the slowly rotating limit (SRL). a) Temperature field image adapted from the $Ra = 10^8$; $Pr = 1$; radius ratio $\chi = 0.6$ spherical shell simulation of \citet{Gastine15}, corresponding to $\Ref = 10^4$.  Lighter (darker) shading represents warmer (cooler) fluid. b) Laboratory shadowgraph image courtesy of Jewel Abbate (UCLA) showing convection in 1.5 cSt silicone oil for $Ra \simeq 4 \times 10^{11}$ and $Pr \simeq 21$, corresponding to $\Ref \simeq 10^5$. The cylindrical tank is 40 cm high by 20 cm across, with shape distorted and its left hand side clipped by the shadowgraph  technique. The horizontal line near the mid-plane and the dark region at the tank bottom are further lighting artifacts.} 
\label{F:RBC}
\end{center}
\end{figure*}

In the limits of asymptotically high $Ra$, high $Re$, turbulent convection, we presume that perfect power law scaling behaviors exist to describe the heat and momentum transport in terms of the other relevant system parameters, $Nu(Ra, Pr)$ and $Re(Ra, Pr)$ \cite[cf.~][]{Grossmann00, Chilla2012, Stevens13, Verma18}. The demonstration of such asymptotic scalings is still an active and frothy topic of scientific debate \cite[e.g.,][]{Glazier99, Lohse03, He12, Lepot18, Doering19, Zhu19}. We assume, further, that similar transport scalings exist in the non-rotating and slowly rotating regimes. Despite, small differences due to symmetry breaking in slowly rotating systems \citep[e.g.,][]{Brown06Coriolis, JQZ17}, their gross transport behaviors can be taken to be comparable (e.g.,  Figure \ref{F:RBC}).

For both non-rotating and slowly rotating convection, we take the characteristic convection length scale to be the global scale of the system in all directions, $\ell \sim H$, based on the superstructures that form at high $Ra$ with vertical scales of order $H$ and lateral scales that are typically less than 10$H$ \citep[e.g.,][]{vonHardenberg08, Adrian16, Pandey18, Vogt18, Akashi19}, which appear to be maintained even in extreme astrophysical and geophysical systems \citep[e.g.,][]{Rieutord2010}. 
In the turbulent limit, the free-fall inertial balance is achieved
\be
\bsu \cdot \nabla \bsu \sim \bsg \alpha  \theta \quad \longrightarrow \quad \frac{U^2}{H} \sim g \alpha \vartheta.
\ee
Analytic estimations for the characteristic magnitude of $\vartheta$ in the turbulent regime are non-trivial \citep[e.g.,][]{Ahlers09, Verma18}. Here, following the work of \citet[e.g.,][]{Grossmann2011}, we scale $\vartheta \sim \Delta T$.
In the NRL and SRL, it then follows that  
\be
\ell \sim H , \,\,   \vartheta \sim \Delta T \mbox{ and } U \sim \sqrt{g \alpha \Delta T H} \equiv \uf \, .
\label{SRLscales}
\ee
The dominant flows in these regimes are large scale; these flows are driven by thermal fluctuations that are roughly comparable to the temperature drop across the system (likely akin to the characteristic boundary layer temperature variations); and the convective flows will approach $\uf$, the diffusivity-free, inertial free-fall velocity \citep[e.g.,][]{Spiegel63, Spiegel71, Chilla2012}. Further, the characteristic advective time scales are isotropic and are given by 
\be
\tau_U = \frac{\ell}{U} \sim \frac{H}{\uf} = \sqrt{H/ (g \alpha \Delta T)} \equiv \tauf \, , 
\ee
where $\tauf$ is the inertial free-fall time across the system. 
We note, following \citet{Spiegel63}, that our assumption that transport processes are dominated by the large scale flows likely best applies in low $Pr$ fluids \citep[e.g.,][]{Vogt18}.  We will not probe this assumption more deeply here, but direct readers to more focussed treatments of non-rotating RBC \citep[e.g.,][]{Grossmann04, Ahlers09, Verma18, Doering20}. 

Using the SRL scales (\ref{SRLscales}), the local Rossby number can be recast as 
\be
\Rol = \frac{U}{2 \Omega \ell}  \sim \frac{\uf}{ 2 \Omega H} = \sqrt{\frac{g \alpha \Delta T}{4 \Omega^2 H}} 
\equiv \RoC \, ,
\label{SRLRol}
\ee
which demonstrates that the local Rossby number is equivalent to the convective Rossby number, $\RoC$, in the slowly rotating limit ($\Rol \gg 1$).  Further, from (\ref{SRLRol}), we arrive at the standard, time-scale based description of the convective Rossby number, $\RoC = \tau_\Omega/\tauf$, as the ratio of the rotational time, $\tau_\Omega = 1/(2 \Omega)$, and the free-fall time across the system-scale. 

The scales in (\ref{SRLscales}) lead to the following NRL and SRL transport estimates
\begin{subequations}
\setlength{\abovedisplayskip}{1pt}
\setlength{\belowdisplayskip}{1pt}
\begin{multline}
Re_\ell = \frac{U \ell}{\nu} \sim \frac{\uf H}{\nu} =  \frac{\sqrt{g \alpha \Delta T H^3} }{\nu} \\ = \left( \frac{Ra}{Pr} \right)^{1/2} \equiv \Ref \, , 
\end{multline}
\bml
Pe_\ell = \frac{U \ell}{\kappa} \sim \frac{\uf H}{\kappa} = \frac{\sqrt{g \alpha \Delta T H^3} }{\kappa} \\ = \left( Ra \, Pr \right)^{1/2} \equiv \Pef \, ,
\end{multline}
\be
Nu \sim \frac{U \vartheta H}{\kappa \Delta T} \sim \frac{\uf H}{\kappa} =  \left( Ra \, Pr \right)^{1/2} \equiv \Pef \, . 
\ee
\label{SRLscalings}
\end{subequations}
Here, $\Ref$ and $\Pef$ are the classic diffusivity-free, free-fall transport parameters. 

Dimensional analysis can be used, independently, to solve for the coefficients $\zeta$ and $\chi$ that yield diffusivity-free expressions for the characteristic transport parameters \citep[e.g.,][]{Plumley19}, yielding
\begin{subequations}
\setlength{\abovedisplayskip}{1pt}
\setlength{\belowdisplayskip}{1pt}
\bml
Re_H \sim \wRa^\zeta Pr^\chi =  \left( Ra/Pr \right)^{1/2} \equiv \Ref \\ (\zeta = - \chi = 1/2), 
\end{multline}
\bml
Pe_H \sim \wRa^\zeta Pr^\chi =  \left( Ra \, Pr \right)^{1/2} \equiv \Pef \, \\ (\zeta = \chi = 1/2), 
\end{multline}
\bml
Nu \sim \wRa^\zeta Pr^\chi =  \left( Ra \, Pr \right)^{1/2} \equiv \Pef \, \\ (\zeta = \chi = 1/2), 
\end{multline}
\label{SRLdiman}
\end{subequations}
where $\wRa \mapsto Ra$ in the dimensional analysis, since $Ra_{crit}$ is effectively constant in the non-rotating and slowly rotating limits. Since it is being assumed that the convection is highly supercritical and turbulence dominated, we take $(Nu -1)  \approx Nu$, $(Re -1)  \approx Re$, and $(\wRa -1) \approx \wRa$ in all our dimensional analyses \citep[cf.][]{Ecke15}. 

The dimensional analytical transport estimates in (\ref{SRLdiman}) are consistent with the dynamical scaling estimates given in (\ref{SRLscalings}), and also agree with the classic dimensional analysis predictions for non-rotating convection  in the limit of zero diffusive effects \citep{Spiegel71}. The agreement between the independent scalings (\ref{SRLscalings}) and (\ref{SRLdiman}) shows that $Re_\ell \sim Re_H$ and $Pe_\ell \sim Pe_H$, consistent with our assumption that $\ell \sim H$ in NRL and SRL. 
Lastly, multiplying by $E$, the momentum transport scalings (\ref{SRLscalings}a) and (\ref{SRLdiman}a) require that 
\be
\Rol \sim Ro_H \sim \RoC
\label{SlowRol}
\ee
in the slowly rotating regime, consistent with (\ref{SRLRol}).

\section{The Rapidly Rotating Limit (RRL)}
\label{sectRRL}
Just as angular momentum is the key dynamical variable in rapidly rotating solid mechanics problems, vorticity, $\bs{\omega} = \nabla \times \bsu$, is the essential dynamical variable in rapidly rotating fluid systems in which rotational inertia dominates the physics \citep{Greenspan68}.  The evolution equation for fluid vorticity, $\nabla \times $ (\ref{GE}a), is:
\bml
\partial_t \bso + \bsu \cdot \nabla \bso - \bso \cdot \nabla \bsu = \\ 2 \bsO \cdot \nabla \bsu  + \nabla \times (\bsg \alpha \theta) + \nu \nabla^2 \bso \, .
\label{VE}
\end{multline}
In the turbulent rapidly rotating limit, a balance is achieved in (\ref{VE}) between the inertial (I), Coriolis (C), and buoyancy (A, for Archimedean) terms \citep{Ingersoll82, Aubert01}.  This is typically referred to as the CIA balance \citep{Jones11, KingBuffett13}:
\begin{gather}
\label{CIAscalingterms}
\bsu \cdot \nabla \bso  \sim 2 \Omega \, \partial_z \bsu  \sim  \nabla \times (\bsg \alpha \theta) \\
\frac{U^2}{\ell^2}  \sim  \frac{2 \Omega U}{H}   \sim  \frac{g\alpha \vartheta}{\ell} \nn
\end{gather}
in which the first term is inertial advection of vorticity (I), the second is the axial stretching of planetary (or background) vorticity (C), and the third is the  buoyancy torque (A).  

\begin{figure*}[t!]
\begin{center}
\centerline{\includegraphics[width=1\textwidth]{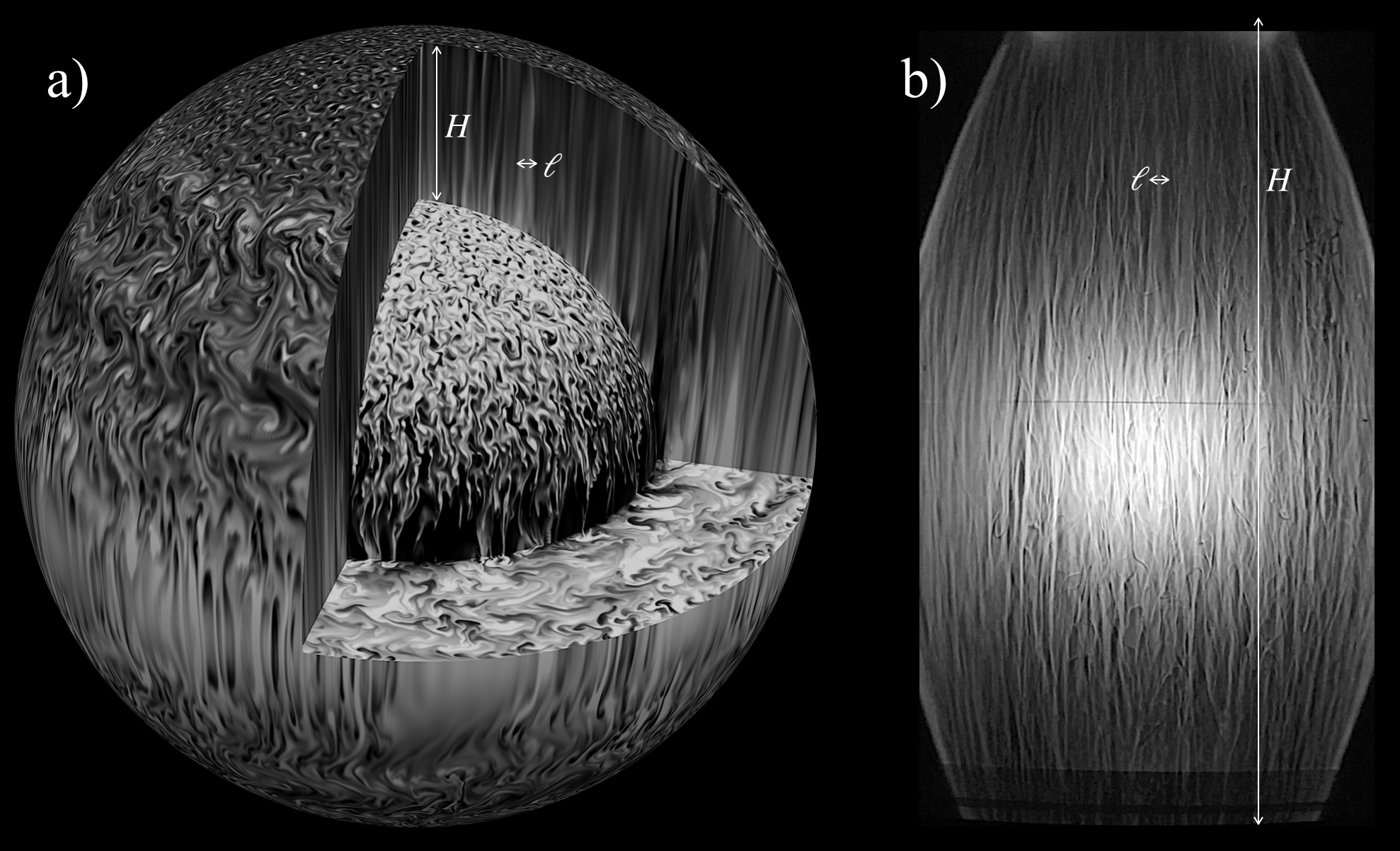}}
\caption{\small Anisotropic flows in rapidly rotating convection with longer characteristic axial scales than horizontal scales $\ell$ (measured perpendicular to the rotation axis). a) Temperature field image from the $Ra = 2.5 \times 10^9$; $E = 10^{-6}$; $Pr = 1$; radius ratio $\chi = 0.6$ spherical shell simulation of \citet{Gastine16}, corresponding to $\RoC = 5 \times 10^{-2}$ and $\ReTW \simeq 1.3 \times 10^2$.   Lighter (darker) shading represents warmer (cooler) fluid. b) Laboratory shadowgraph image courtesy of Jewel Abbate (UCLA) showing rotating convection in 1.5 cSt silicone oil for $Ra \simeq 5 \times 10^{11}$; $E \simeq 6 \times 10^{-7}$; $Pr \simeq 21$, corresponding to $\RoC \simeq 9 \times 10^{-2}$ and $\ReTW \simeq 1.3 \times 10^3$. The cylindrical tank is 40 cm high by 20 cm across, with its shape distorted and clipped around the mid-plane by the shadowgraph imaging technique. The horizontal line near the mid-plane and the dark region at the tank bottom are further lighting artifacts.} 
\label{F:Aniso}
\end{center}
\end{figure*}

Rapidly rotating convective motions are strongly anisotropic, as shown in Figure \ref{F:Aniso}, with small scales perpendicular to ${\bsO}$ and much longer scales parallel to ${\bsO}$.  Therefore, it is essential in (\ref{CIAscalingterms}) to distinguish between the characteristic convection scale $\ell$ measured perpendicular to ${\bsO}$ and the system scale $H$ measured parallel ${\bsO}$. Only the stretching of the background vorticity, $2 \Omega \, \partial_z \bsu$, can occur on the system scale.  The other two terms, I and A, operate on the local convective scale. Although the length scales $\ell$ and $H$ differ greatly in rapidly rotating convection, the kinetic energies measured along these different directions remain comparable, even in the supercritical regime \citep{Stellmach04, Sprague06, Julien12gafd, Horn15, Vogt20}. Thus, we assume that the characteristic velocity magnitudes are approximately isotropic $|u_i| \sim U$ in RRL. 

The balance between the C and I terms in (\ref{CIAscalingterms}) then gives 
\be
\frac{\ell}{H} \sim \frac{U}{2 \Omega \ell} \equiv \Rol = \frac{\tau_\Omega}{\tau_U^\ell} \, ,
\label{CI}
\ee
where the lateral advective time scale $\tau_U^\ell = \ell/U$ characterizes rapidly rotating convection. 
Thus, rapidly rotating convection is highly anisotropic with $\ell \ll H$, since $\ell \sim \Rol H$ in (\ref{CI}) and $\Rol \ll 1$ in the definition of the RRL. 
Unlike in the NRL and SRL where the bulk fluid tends to be isothermalized by strong turbulence, in rapidly rotating convection, an unstable mean temperature gradient tends to be sustained in the fluid bulk, $\partial_g \overline{T} \sim \Delta T/H$ \citep[e.g.,][]{Julien96, Hart99, Julien12gafd, Cheng20}. The fluctuating thermal energy equation (\ref{GE}b) thus scales as
\be
(\bsu \cdot \nabla \theta)^\prime \sim (\bs{\hat{e}_g} \cdot \bsu) \partial_g \overline{T} \,\, \longrightarrow \,\, \frac{U \vartheta}{\ell} \sim \frac{U \Delta T}{H}.
\ee
This implies, in the RRL, that 
\be
\frac{\vartheta}{\Delta T} \sim \frac{\ell}{H} \sim \Rol \, .
\ee

Balancing the C and A terms in (\ref{CIAscalingterms}) yields
\be
U 
\sim \frac{g \alpha \Delta T}{2 \Omega} \left( \frac{\vartheta}{\Delta T} \, \frac{H}{ \ell} \right) \sim \frac{g \alpha \Delta T}{2 \Omega} \equiv \uTW \, ,
\label{CAscalingterms}
\ee
where $\uTW$, the thermal wind velocity, is the diffusivity-free velocity scale in the rapidly rotating convection regime \citep{McWilliams06}.  (This thermal wind scaling is similarly found by balancing the I and A terms in (\ref{CIAscalingterms}).) From (\ref{CAscalingterms}), we see that the local advection time scale in RRL is the thermal wind time scale: 
\be
\tau_U^\ell = \ell / \uTW \equiv \tau_{T\!W} \, .
\ee
The rapidly rotating local Rossby number then becomes
\bml
\Rol \sim \frac{\uTW}{2 \Omega \ell} 
=\frac{\tau_\Omega}{\tauTW}
= \frac{g \alpha \Delta T}{(2 \Omega)^2 H} \frac{1}{\Rol} \\ \,\, \longrightarrow \,\, 
\Rol \sim \sqrt{ \frac{Ra E^2}{Pr}}  \equiv \RoC \, .
\label{RRLRol}
\end{multline}
Thus, the {\it a posteriori} local Rossby number, $\Rol$, is equivalent to the {\it a priori} convective Rossby number, $\RoC$, in both the slowly rotating limit  (\ref{SRLRol}) and in the rapidly rotating limit (\ref{RRLRol}). At closer inspection, this holds because the local advective time scales, $\tauf = H/\uf$ in SRL and $\tauTW = \ell / \uTW$ in RRL, are similar.  Thus, their ratio yields
\be
\frac{\tauf}{\tauTW} \sim \frac{H}{\ell} 
\, \frac{\uTW}{\uf} \sim \frac{1}{\RoC} 
\, \frac{g \alpha \Delta T / (2 \Omega)}{\sqrt{g \alpha \Delta T H}} = \mathcal{O}(1) \, .
\label{sameTimes}
\ee
This similarity between the SRL and the RRL local advective time scales explains why the convective Rossby number turns up so ubiquitously in rotating convection dynamics: even though $\uf$ and $H$ in SRL both greatly exceed $\uTW$ and $\ell$ in RRL, their ratios, $\uf / H$ and $\uTW/\ell$ have equivalent scaled values. Expression (\ref{sameTimes}) demonstrates, further, that the convective Rossby number can be cast, alternatively, as 
\be
\RoC \equiv \frac{\uTW}{\uf} \, .
\label{RoC-VelRatio}
\ee
This velocity-based definition of $\RoC$ holds in both slowly rotating and rapidly rotating regimes, and differs in its interpretation in comparison to the standard (slowly rotating) definition in which $\RoC = \tau_\Omega / \tauf$, as will be discussed further in section \ref{Interp}. 

In the limit of rapidly rotating convective turbulence, the CIA balance gives
\begin{gather}
\ell \sim \RoC H \, , \quad  \vartheta \sim \RoC \Delta T \, , \\  U \sim \RoC \uf =  \frac{g \alpha \Delta T}{2 \Omega} \equiv \uTW \, ,
\label{RRLscales}
\end{gather}
with all three turbulent RRL scales differing by $\RoC$ relative to their corresponding SRL scales. 
Following the same steps as in (\ref{SRLscales}) but employing the rapidly rotating scales in (\ref{RRLscales}) then leads to the following RRL transport estimates
\begin{subequations}
\setlength{\abovedisplayskip}{1pt}
\setlength{\belowdisplayskip}{1pt}
\bml
Re_\ell = \frac{\uTW \ell}{\nu} \sim \frac{g \alpha \Delta T}{2 \Omega}\frac{\ell}{\nu} 
= \left(\frac{Ra}{Pr} \right)^{3/2} E^2  \\ =  \RoC^2 \Ref  \equiv \ReTW \, ,
\end{multline}
\bml
Pe_\ell = \frac{\uTW \ell}{\kappa} \sim \frac{g \alpha \Delta T}{2 \Omega} \frac{\ell}{\kappa} 
= \left(\frac{Ra^{3/2}}{Pr^{1/2}} \right) E^2 \\ =  \RoC^2 \Pef  \equiv \PeTW \, , 
\end{multline}
\bml
Nu \sim \frac{\uTW }{\kappa} \left( \frac{\vartheta H}{\Delta T} \right)    = \frac{\uTW \ell}{\kappa}
= \left(\frac{Ra^{3/2}}{Pr^{1/2}} \right) E^2  \\ =  \RoC^2 \Pef  
\equiv Pe_{TW} \, , 
\end{multline}
\label{RRLscalings}
\end{subequations}
where $\ReTW$ and $\PeTW$ are the thermal wind Reynolds and thermal wind P\'eclet numbers, respectively. 

The scaling analysis in (\ref{RRLscalings}) is consistent with rapidly rotating, diffusivity-free dimensional analysis, which yields
\begin{subequations}
\setlength{\abovedisplayskip}{1pt}
\setlength{\belowdisplayskip}{1pt}
\bml
Re_\ell \sim \wRa^\zeta Pr^\chi =  \left( Ra/Pr \right)^{3/2} E^2  \equiv  \ReTW \\ (\zeta = - \chi = 3/2) \, , 
\end{multline}
\bml
Pe_\ell \sim \wRa^\zeta Pr^\chi = \left(\frac{Ra^{3/2}}{Pr^{1/2}} \right) E^2   \equiv  \PeTW \\ (\zeta = -3 \chi = 3/2) \, ,  
\end{multline}
\bml
Nu \sim \wRa^\zeta Pr^\chi =  \left(\frac{Ra^{3/2}}{Pr^{1/2}} \right) E^2  \equiv   \PeTW \\ (\zeta = -3 \chi = 3/2) \, .
\end{multline}
\label{RRLdiman}
\end{subequations}
Note that the critical Rayleigh number in (\ref{RRLdiman}) varies strongly with the system's rapid rotation, $\wRa \sim Ra E^{4/3}$. Consistency between (\ref{RRLscalings}) and (\ref{RRLdiman}) requires that the pertinent velocity and length scales must be $\uTW$ and $\ell$ in RRL. Thus, $Re \sim Re_\ell \equiv \ReTW$ and $Pe \sim Pe_\ell \equiv \PeTW$ in the rapidly rotating regime. Multiplying (\ref{RRLscalings}a) by the local Ekman number, $E_\ell = \nu/(2\Omega \ell^2)$, yields $\Rol = Re_\ell E_\ell \sim \RoC$, consistent with (\ref{RRLRol}). 
%
%
Further, the RRL heat transport scaling (\ref{RRLscalings}c) is also consistent with asympotically reduced theory and diffusivity-free formulations \citep[e.g.,][]{Julien12, Barker14, Plumley19, Maffei20}.  
Recent studies, such as \citet{Plumley16, Plumley17}, suggest that it is possible to reach the RRL scalings (\ref{RRLscalings}) at far lower $\widetilde{Ra}$ values than are necessary to reach their diffusivity-free non-rotating counterparts \citep[cf.][]{Doering20}.

The rapidly rotating thermal wind transport scalings in (\ref{RRLscalings}) differ from the slowly rotating free-fall scalings by a factor of $\RoC^2$.  This creates a clean and novel link between the two sets of scaling predictions.  We can alternatively cast the RRL expressions as
\begin{gather}
\label{RoCubed}
Re_\ell \sim \RoC^3 \, E^{-1} , \,\, \mbox{and}  \nn \\
Pe_\ell \sim Nu \sim \RoC^3 \, (E/Pr)^{-1} \, .
\end{gather}
From (\ref{CIAscalingterms}), we predict that rapidly rotating turbulent transport data acquired with approximately fixed rotation rate and material properties will be collapsed when normalized by the cube of the convective Rossby number.

Local scale parameterizations naturally arise in our analysis of rapidly rotating transport phenomena. However, the system scale transport parameters, $Re_H$ and $Pe_H$, are most often reported in the literature \citep[e.g,][]{Guervilly16}. 
Thus, we rescale our local rapidly rotating transport scalings to provide the equivalent, system scale counterparts: 
\begin{subequations}
\setlength{\abovedisplayskip}{1pt}
\setlength{\belowdisplayskip}{1pt}
\bml
Re_H = Re_\ell \, \frac{H} {\ell} \sim \frac{\uTW H}{\nu} = \frac{Ra E}{Pr} \\
           =  \RoC^{-1} \, \ReTW   = \RoC \Ref  
\label{ReH}
\end{multline}
\bml
Pe_H = Pe_\ell \, \frac{H} {\ell} \sim \frac{\uTW H}{\kappa}  = Ra E \\
	= \RoC^{-1} \, \PeTW   = \RoC \Pef . 
\end{multline}
\label{SysScales}
\end{subequations}
In addition, the system scale Rossby number scales as   
\be
Ro_H = Re_H E \sim \RoC^2 \, , 
\label{RoSquared}
\ee
in agreement with the low-$E$, quasi-geostrophic convection models \citet{Guervilly19} and the three-dimensional asymptotically-reduced models of \citet{Maffei20}. This system scale RRL Rossby number scaling (\ref{RoSquared}) differs by a factor of $\RoC$ relative to the slowly rotating scaling (\ref{SlowRol}) in which $Ro_H \sim \Rol \sim \RoC$.

\section{Flux-Based Scalings}
\label{sectFlux}

\subsection*{Non-Rotating and Slowly Rotating Flux-Based Scalings}
When considering a planetary or stellar convection system, it is far easier to estimate the outward thermal flux than to infer a temperature drop across a given fluid layer. Therefore, it is of great utility to recast the 
scalings developed above in terms of the (superadiabatic) heat flux, $Q$, instead of the temperature difference, $\Delta T$. 
Non-dimensionally, this simply corresponds to replacing the Rayleigh number, $Ra \propto \Delta T$, with the flux Rayleigh number, $Ra_F = Ra Nu \propto Q$.  In order to recast the NRL and SRL scalings in terms of $Ra_F$, we manipulate (\ref{SRLscalings}c) into the form
\be
Ra \sim \left[ Ra Nu Pr^{-1/2} \right]^{2/3}     \sim  Ra_F^{2/3}  Pr^{-1/3}  \, , 
\label{SRLfluxHX}
\ee
and substitute this into (\ref{SRLscalings}a) and (\ref{SRLscalings}b), giving the flux-based free-fall scalings
\bea
\Ref \sim \left[ \frac{Ra_F}{Pr^2} \right]^{1/3}  \,\, \mbox{and} \,\,
\Pef  \sim \left[ Ra_F \, Pr \right]^{1/3} .
\eea
The SRL flux-based expression for the Rossby number is then \citep[cf.][]{Maxworthy94}
\be
Ro_\ell \sim Ro_H 
\sim \left[ \frac{Ra_F E^3}{Pr^2} \right]^{1/3} \, = {Ra_F^*}^{1/3} .
\label{FluxRoSRL}
\ee

The respective dimensional forms of the uncontrolled temperature drop (which is assumed here to be proportional to $\vartheta$ in NRL and SRL) and the free-fall velocity scale in the slowly rotating regime are 
\begin{subequations}
\setlength{\abovedisplayskip}{1pt}
\setlength{\belowdisplayskip}{1pt}
\bea
\Delta T \sim Q / \rho c_P \uf  \sim   \left (  \frac{Q^2}{g \alpha \rho^2 c_P^2 H} \right )^{1/3} \, , \\
U \sim \uf \sim \sqrt{g \alpha \Delta T H} \sim \, \left[  \frac{g \alpha Q H}{\rho c_P} \right]^{1/3} .
\eea
\label{SRLqtot}
\end{subequations}
Equations (\ref{SRLqtot}) correspond to the free-fall balance 
expressed in terms of an applied heat flux $Q$ \citep[e.g.,][]{Deardorff70}.
Further, by inserting (\ref{SRLqtot}a) into (\ref{RoCdefn}), we find that the flux-based SRL expression for $\RoC = {Ra_F^*}^{1/3}$, identical to $\Rol$ in (\ref{FluxRoSRL}). Thus, $\Rol \approx \RoC$ in the flux-based framework as well, as must be the case since this result is framework independent. 

\subsection*{Rapidly Rotating Flux-Based Scalings}
In order to formulate the flux-based, system-scale, rapidly rotating momentum transport scaling, we recast the RRL heat transport scaling (\ref{RRLscalings}c) as 
\begin{align}
Ra & = (Ra Nu Pr^{1/2} E^{-2})^{2/5} \nn \\
     & = Ra_F^{2/5} Pr^{1/5} E^{-4/5} \, .
\label{RRLfluxHX}
\end{align}
%
Substituting (\ref{RRLfluxHX}) into (\ref{RRLscalings}) leads to the local, flux-based, rapidly rotating transport scalings:
\begin{subequations}
\setlength{\abovedisplayskip}{1pt}
\setlength{\belowdisplayskip}{1pt}
\label{fluxlocal}
\bml
Re_{TW}  \sim \left[ \frac{Ra_F E^{4/3}}{Pr^2} \right]^{3/5}  \\  = {Ra_F^*}^{3/5} E^{-1} = \RoC^3 E^{-1} \, ,
\end{multline}
\bml
Pe_{TW}  \sim \left[ \frac{Ra_F E^{4/3}}{Pr^{1/3}} \right]^{3/5}  \\ = {Ra_F^*}^{3/5} (E/Pr)^{-1} = \RoC^3 (E/Pr)^{-1} \, ,
\end{multline}
\end{subequations}
with the local scale, RRL flux-based Rossby number 
\be
\Rol \sim \left[ \frac{Ra_F E^3}{Pr^2} \right]^{1/5}  \,  = {Ra_F^*}^{1/5} = \RoC.
\label{FluxRoRRL} 
\ee
Note, using the flux-based expression for $U_{TW}$ (given in (\ref{RRLdimeqtns}) below), that again $\Rol \approx \RoC$. However, in contrast to the fixed temperature configuration, the flux-based local Rossby numbers in the SRL  (\ref{FluxRoSRL}) and the RRL (\ref{FluxRoRRL}) are no longer identical. Instead, both flux-based $\Rol$ expressions depend on the modified flux Rayleigh number, but in the SRL $Ra_F^*$ is raised to the one-third power, whereas it is raised to the one-fifth power in the RRL.  This difference in the flux-based $\Rol$ expressions stems from the different $Ra(Ra_F, E, Pr)$ scalings given in (\ref{SRLfluxHX}) and (\ref{RRLfluxHX}). 

The system scale, flux-based, rapidly rotating transport scalings are often used in the geophysical and astrophysical literature \citep[e.g.,][]{Gillet06, Christensen10}.  These are found by substituting (\ref{RRLfluxHX}) into (\ref{ReH}), which leads to 
\begin{subequations}
\begin{align}
\label{CIA}
Re_H  \sim \left[ \frac{Ra_F}{Pr^2} \right]^{2/5} E^{1/5} 
& \equiv Re_{C\!I\!A}  \, , \\
Pe_H  \sim Ra_F^{2/5} (E \, Pr)^{1/5}  & = Re_{C\!I\!A} Pr \, .
\end{align}
\end{subequations}
The flux-based $Re_H$ expression (\ref{CIA}) is referred to as the CIA scaling velocity, $Re_{C\!I\!A}$, since it is indeed derived from the CIA triple balance \citep[e.g.,][]{Ingersoll82, Aubert01, Jones11, KingBuffett13}. This flux-based momentum transport scaling is easily converted back into a temperature-based scaling by substituting $Ra_F = Ra Nu$ into $Re_{C\!I\!A}$ and then further substituting $Nu \sim Ra^{3/2} E^2 / Pr^{1/2} = \PeTW$. Doing so yields
\be
Re_{C \! I \! A} \sim \left[ \frac{Ra}{Pr^2}  \left( \frac{Ra^{3/2} E^2}{Pr^{1/2}} \right) \right]^{2/5}  E^{1/5}   
=  \RoC \Ref \, , 
\label{KJarg}
\ee
in agreement with (\ref{RRLscalings}a) and (\ref{ReH}). Multiplying (\ref{KJarg}) by $E$ then demonstrates that 
\be
Ro_{C \! I \! A} \sim \RoC^2 \quad \,\, (\mbox{when} \,\,\, Nu \rightarrow \PeTW).
\label{CIAasymptote}
\ee
Thus, 
the classical, flux-based CIA theory is synonymous with the temperature-based rapidly rotating velocity scalings given in (\ref{SysScales}a) and (\ref{RoSquared}).

Since most laboratory and numerical simulations cannot reach the diffusivity-free $\PeTW$ heat transfer trend, the $Ro_H \sim \RoC^2$ scaling is difficult to attain \citep[cf.][]{Bouillaut19, Guervilly19, Maffei20}. For example, in the seminal planetary dynamo survey of \citet{Christensen06}, it was found that $Ro_H \sim {Ra_F^*}^{2/5}$, which, comparing to (\ref{CIA}), shows that the bulk flow had attained the turbulent, CIA scaling.  Their heat transfer data was best fit as $Nu \sim Ra E$, which differs from the $\PeTW$ scaling likely because it was controlled by diffusive, boundary layer physics \citep[e.g.,][]{Cheng16, Julien16, Hawkins20}. This corresponds to $Ro_H \sim \RoC^{8/5}$. However, if we substitute $Nu = Ra^{3/2} E^2/Pr^{1/2}$ in place of their $Nu \sim Ra E$ scaling, then the system-scale Rossby number scaling necessarily transforms to $Ro_H \sim \RoC^2$.

Our flux-based momentum transport scalings show that the RRL transport (\ref{ReH}) is formally identical to the classical, flux-based CIA velocity scaling (\ref{CIA}) when $Nu \approx Pe_{TW}$. However, this $Pe_{TW}$ heat transfer scaling is not often found in standard experiments or direct numerical simulations, because the heat transfer rarely reaches the RRL trend \citep[cf.][]{Julien12, Barker14, Plumley16, Plumley17}. This is an important physical point, as the flux-based $Re_H$ scaling in (\ref{CIA}) can be applied for any $Nu$ value and, accordingly, is often considered to be fundamentally different from, and to conflict with, the local scale prediction (\ref{RRLscalings}a) and the system scale prediction (\ref{ReH}) that both naturally arise in the $Nu \approx Pe_{TW}$ rapidly rotating scaling turbulent arguments given here and in rapidly-rotating asymptotic analysis \citep{Sprague06, Calkins15, Calkins18, Maffei20}. Directly comparing the Reynolds numbers scalings in (\ref{RRLscalings}a) and (\ref{CIA}) is, however, incorrect since they are defined on different length scales.  In contrast, it is appropriate to compare (\ref{ReH}) and (\ref{RRLscalings}a) since they are both system-scale quantities, and we have shown, in fact, that these scalings are identical in the turbulent rapidly rotating limit where $Nu \rightarrow Pe_{TW}$.

The respective dimensional forms of the rapidly rotating temperature fluctuation, temperature drop, and velocity scales are: 
\begin{subequations}
\setlength{\abovedisplayskip}{1pt}
\setlength{\belowdisplayskip}{1pt}
\label{RRLdimeqtns}
\bea
 \vartheta &\sim& \frac{Q}{\rho c_P \uTW}  \sim 
  \left ( \frac{2\Omega}{g^2 \alpha^2 H} \right )^{1/5}   \left (\frac{Q}{\rho c_p} \right )^{3/5}  \, , \\
 \Delta T &\sim& \frac{\vartheta H}{\ell} \sim \frac{\left( 2 \Omega\right )^{4/5} H^{1/5}}{\left ( g\alpha \right )^{3/5}}
  \left (\frac{Q}{\rho c_p} \right )^{2/5} 
  \ , 
 \eea
 \vspace{-0.35cm}
 \bml
 U \sim U_{TW} \sim \sqrt{g\alpha \vartheta \ell} \sim \frac{g \alpha \vartheta}{2 \Omega} \frac{H}{\ell} \\ \sim
 \left (\frac{g\alpha Q}{\rho c_p} \right )^{2/5} \left ( \frac{H}{2\Omega} \right )^{1/5} \, .
\end{multline}
\end{subequations}

In this section, we have transformed the scaling results produced in the $\Delta T$-based framework to the $Q$-based framework through the definition of the flux Rayleigh number $Ra_F = Ra Nu$. However, in the flux-based scalings, we find a lack of equivalence between the SRL and RRL local Rossby numbers, thereby preventing us from making simple, flux-based connections between the scaling regimes, as is possible in the temperature-based framework. Nevertheless, exploration of the flux-based framework has shown that the classical, flux-based CIA scalings produced in many prior works is formally synonymous with the temperature-based scalings developed herein (cf.~(\ref{RoSquared}) and (\ref{CIAasymptote})).

\begin{table*}[t!]
\begin{center}
\resizebox{\textwidth}{!}{
\begin{tabular}{|c|c|c|c|c|c|c|c|c|c|c|}
\hline
Regime 	& $\Rol  $ & $\ell $ & ${\vartheta} $ & $U$ 	& $Re_\ell$ & $Pe_\ell$ & $Nu $ & $Re_H $ & $Pe_H $ & $Ro_H $  \\
($\Delta T$-based)	& $\approx \! \RoC $ &  & &  	&  &  &  &  &  &   \\
\hline 
SRL 		& $\gg 1$ &  $H$ & $ \Delta T$& $\uf$ & $\Ref $ & $\Pef $ & $\Pef$ & $\Ref $  & $\Pef $ & $\RoC $\\ 
\hline
RRL		& $\ll 1$   &  $\RoC H$ & $\RoC \Delta T$ & $\RoC \uf$ & $\RoC^2 \Ref $ & $\RoC^2 \Pef $ & $\RoC^2 \Pef $ & $\RoC \Ref$ & $\RoC \Ref$ & $\RoC^2$ \\
\hline	
\end{tabular}
} 
\end{center}
\caption{\small Summary of applied $\Delta T$, turbulent scaling estimates for characteristic convective scales and transports in the slowly rotating limit (SRL) and the rapidly rotating limit (RRL). The free fall velocity is defined here as $\uf \sim \sqrt{g \alpha \Delta T H}$ and $\Rol \sim \RoC = \sqrt{Ra E^2/Pr}$ in both SRL and RRL. The non-rotating (NRL) scalings are identical to SRL in our treatment, excepting that the Rossby number is not defined in the non-rotating regime.} 
\label{T:Sum}
\end{table*}%

\begin{table*}[t!]
\begin{center}
\resizebox{\textwidth}{!}{
\begin{tabular}{|c|c|c|c|c|c|c|c|c|c|c|}
\hline
Regime 	& $\Rol  $ & $\ell $ & ${\vartheta} $ & $U$ 	& $Re_\ell$ & $Pe_\ell$ & $Nu $ & $Re_H $ & $Pe_H $ & $Ro_H $  \\
($Q$-based)	& $\approx \! \RoC $ &  & &  	&  &  &  &  &  &   \\
\hline 
SRL 		& $\gg 1$ &  $H$ & $ \Delta T$& $\uf$ & $\Ref $ & $\Pef $ & $\Pef$ & $\Ref $  & $\Pef $ & $\RoC $\\ 
\hline
RRL		& $\ll 1$   &  $\RoC H$ & $\RoC \Delta T$ & $\RoC^{1/3} \uf$ & $\RoC^{4/3} \Ref $ & $\RoC^{4/3} \Pef $ & $\RoC^{4/3} \Pef $ & $\RoC^{1/3} \Ref$ & $\RoC^{1/3} \Ref$ & $\RoC^2$ \\
\hline	
\end{tabular}
} 
\end{center}
\caption{\small Summary of applied $Q$, turbulent scaling estimates for characteristic convective scales and transports in the slowly rotating limit (SRL) and the rapidly rotating limit (RRL). The free fall velocity is defined here as $\uf \sim (g \alpha Q H / \rho c_P)^{1/3} = (\mathcal{B}H)^{1/3}$. Note in the flux based framework that $\Rol \sim \RoC \sim {Ra_F^*}^{1/3}$ in the SRL, whereas $\Rol \sim \RoC \sim {Ra_F^*}^{1/5}$ in the RRL. For example, we find $Ro_H \sim \RoC^{2} \sim {Ra_F^*}^{2/5}$ in the rapidly rotating regime, which is consistent with $\Delta T \sim Q^{2/5}$ in (\ref{RRLdimeqtns}b).} 
\label{T:SumQ}
\end{table*}%

\section{Discussion}
\label{Interp}

The convective scaling relationships presented here are generated via exactly parallel constructions, first made within the non-rotating and slowly rotating limits and then secondarily made within the rapidly rotating limit. Starting from the generic non-dimensional transport parameters, $Re = U\ell/\nu$, $Pe = U\ell / \kappa$ and $Nu = U \vartheta / (\kappa \Delta T/H)$, we select the dynamically relevant estimates for $\ell$, $\vartheta$ and $U$ that characterize a given convection system. 
Two configurations of thermal driving are considered: the fixed-temperature regime (Table \ref{T:Sum}), popular for its ease of application and interpretation in modeling studies, and the fixed heat flux regime (Table \ref{T:SumQ}), popular for its ease of application in geophysical and astrophysical settings. 

The fixed-temperature configuration is particularly elegant, and we will focus on the fixed temperature scalings in this discussion. First, our analyses show that the local Rossby number is equivalent to the convective Rossby number, 
\bdm
\Rol \simeq \RoC,
\edm
in both the slowly and the rapidly rotating frameworks, where $\Rol \equiv U / (2 \Omega \ell)$ is estimated using the characteristic convective length $\ell$, the velocity scale $U$ for each limit, and $\RoC \equiv \sqrt{Ra E^2 Pr^{-1}}$.

Second, by taking the ratios of the rapidly rotating and slowly rotating characteristic scales, we find that they are all related via powers of $\RoC^1$,
\be
\frac{\ell}{H} \sim \frac{\vartheta}{\Delta T} \sim \frac{\uTW}{\uf} \sim \RoC \, , 
\ee
Thirdly, we have shown that the RRL thermal wind transports and the SRL free-fall transports differ from one another via powers of $\RoC^2$, 
\be
\frac{\ReTW}{\Ref} \sim \frac{\PeTW}{\Pef} \sim \RoC^2 \, .
\ee
Further, our generic scalings predict that the system-scale Rossby number, $Ro_H$, scales as $\RoC$ in the slowly rotating regime and as $\RoC^2$ in the rapidly rotating regime. Thus, the convective Rossby number is shown to explain the local-scale convection dynamics, $\Rol \approx \RoC$, and is essential for relating the slowly rotating convection behaviors to those of the rapidly rotating regime.  Thus, $\RoC$ (and synonymously $\Rol$) arise ubiquitously in describing rotating convective flows.  Furthermore, the theoretical framework we have developed here provides a novel, and remarkably straightforward, set of experimentally testable interconnections between the slowly rotating and rapidly rotating convective regimes. As summarized in Table 1, these scalings all depend rather simply on $\Rol \approx \RoC$.

We have shown that when $\Rol$ is defined using the appropriate slowly rotating characteristic scales is equivalent to the convective Rossby number $\RoC$: 
\be
\Rol = \frac{\uf}{2 \Omega H} = \frac{\tau_\Omega}{\tauf} = \sqrt{\frac{Ra E^2}{Pr}} \equiv \RoC  \,\, \mbox{(SRL).}
\label{blah!}
\ee
Following from this, $\RoC$ is often interpreted as the ratio between freely falling convective inertia and the system's rotational inertia \citep[e.g.,][]{Gilman77, Julien96, Aurnou07, Horn14}.  This interpretation is accurate in the slowly rotating regime \citep[e.g.,][]{Brown06Coriolis, Gastine13, JQZ17}. In contrast, this $\uf$-based interpretation is not accurate in rapidly rotating cases, where the length and velocities scales are far smaller than in the slowly rotating regime (Table 1). 

Surprisingly, though, we have shown that the $\Rol$ also scales equivalently to $\RoC$ in the rapidly rotating limit: 
\be
\Rol = \frac{\uTW}{2 \Omega \ell} = \frac{\tau_\Omega}{\tauTW} = \sqrt{\frac{Ra E^2}{Pr}} \equiv \RoC  \,\, \mbox{(RRL).}
\ee
This equivalence holds since the free-fall time scale in the slowly rotating regime scales similarly to the thermal wind time scale in the rapidly rotating regime, 
\be
\tauf = \frac{H}{\uf} \sim \frac{\ell}{\uTW} = \tauTW. 
\ee
Thus, the Rossby number based on the dominant dynamical scale is equivalent to the convective Rossby number in both end member rotational regimes, $\Rol \simeq \RoC$.  This makes clear that the convective Rossby number is, in fact, an appropriate descriptor of rapidly rotating convection dynamics, but it should always be cast as $\RoC = \uTW/(2 \Omega \ell)$ in the rapidly rotating limit.  
Further, since $\Rol \simeq \RoC$ in both regimes, $\RoC$ can be further interpreted as the descriptor of the local scale rotating convection dynamics, irrespective of its value. We conclude then that the convective Rossby number really ties the room together.

The fixed heat flux configuration can be deduced from the fixed-temperature configuration through the relation 
$Ra_F = Ra Nu$. We again find that $\Rol \simeq \RoC$ in both the slow rotating and rapidly rotating limits. However, they no longer have a common definition: $\RoC \sim Ra_F^{*1/3}$ in the SRL regime and $\RoC \sim Ra_F^{*1/5}$ in the RRL regime. 
The relationships between the various flux-based scalings are given in Table 2.

Irrespective of the configuration, a clear  interpretation of $\RoC$ arises from our scaling analyses. The two characteristic velocities in rotating convection are $\uf$ and $\uTW$. In slowly rotating convection, $U \sim \uf \ll \uTW$, since all the fluid's buoyant potential energy is converted to kinetic energy well before it reaches $\uTW$. (Alternatively stated, $\uTW$ becomes singularly large as $\Omega$ becomes small.) In rapidly rotating convection, $U \sim \uTW \ll \uf$ since the vortex stretching term in (\ref{VE}) greatly limits the distance through which a rotating parcel of buoyant fluid can actually freely fall \citep{Jones11}. The selection between $\uf$ and $\uTW$ is based on the more restrictive value between the two: 
\be
U \simeq \mbox{min}(\uf, \uTW) \, .
\ee
Since $\RoC = \uTW/\uf$, it can be validly interpreted as the essential control parameter that picks between the two characteristic velocitites: 
\begin{subequations}
\setlength{\abovedisplayskip}{1pt}
\setlength{\belowdisplayskip}{1pt}
\bea
\RoC \gg 1 &\Rightarrow&  \mbox{min}(\uf, \uTW) = \uf , \\
\RoC \sim 1 &\Rightarrow& \mbox{min}(\uf, \uTW)  = U , \\
\RoC \ll 1   &\Rightarrow&  \mbox{min}(\uf, \uTW) = \uTW .
\eea
\label{RoC-U}
\end{subequations}
The relative ordering of the characteristic time scales is also, therefore, set by $\RoC$: 
\begin{subequations}
\setlength{\abovedisplayskip}{1pt}
\setlength{\belowdisplayskip}{1pt}
\bea
\RoC \gg 1  &\Rightarrow&  \tau_\Omega \gg (\tauf \sim \tau_U^H) , \\
\RoC \sim 1 &\Rightarrow& \tau_\Omega \sim \tauf \sim \tau_U^\ell \sim \tau_U^H  ,  \\
\RoC \ll 1  &\Rightarrow& \tau_\Omega \ll (\tauf \sim \tau_U^\ell) \ll \tau_U^H  . 
\eea
\end{subequations}
The intermediate $\RoC \sim 1$ regime has not been considered here.  There is, however, a great deal of laboratory \citep[e.g.,][]{Rossby69, Ecke97, Kunnen08, King09, Weiss11, Ecke15, Cheng15} and numerical simulation data \citep[e.g.,][]{Gilman77, Julien96, Christensen06, Aurnou07, Soderlund12, Gastine13, Gastine13MNRAS, Gastine16, Featherstone15, Mabuchi15, Anders19} in the $\RoC = \mathcal{O}(1)$ regime. Thus, its scaling behaviors are of broad interest and should be considered in future studies.

An array of new convection and rotating convection devices have been recently built at research centers world-wide \citep[e.g.,][]{Cheng18, Lepot18, Zurner19}.  These next-generation laboratory devices, and associated state of the art numerical simulations, will allow investigations into the efficacy and applicability ranges of the turbulent scaling predictions presented here (Tables \ref{T:Sum} and \ref{T:SumQ}). Our goal will then be to test, possibly validate, and disambiguate between these differing scaling laws given high fidelity measurements, and thereby deduce accurate, robust relations for nonrotating, slowly rotating, and rapidly rotating convective heat and momentum transport, as is necessary to explain and interpret industrial, astrophysical and geophysical convection phenomena.

\subsection*{Acknowledgements}  This work arose from discussions at the ``Rotating Convection: From the Lab to the Stars'' workshop held at the Lorentz Center (\url{https://www.lorentzcenter.nl}) in May 2018.  We gratefully acknowledge the financial support of the NSF Geophysics Program (EAR awards 1620649 and 1853196) and  NSF Applied Mathematics Program (DMS award 2009319), NSF Astronomical Sciences (AST award 1821988), NASA (award 80NSS18K1125) and the German Research Foundation (DFG award HO 5890/1-1). Further thanks are given to Jewel Abbate (UCLA) and Thomas Gastine (IPGP) for supplying the images used in Figures \ref{F:RBC} and \ref{F:Aniso}.


\begin{thebibliography}{}

\end{thebibliography}


\begin{thebibliography}{}

\bibitem[Ahlers et~al., 2009]{Ahlers09}
Ahlers, G., Grossmann, S., and Lohse, D. (2009).
\newblock {Heat transfer and large scale dynamics in turbulent
  {R}ayleigh-{B}{\'e}nard convection}.
\newblock {\em Rev.~Mod.~Phys.}, 81(2):503.

\bibitem[Akashi et~al., 2019]{Akashi19}
Akashi, M., Yanagisawa, T., Tasaka, Y., Vogt, T., Murai, Y., and Eckert, S.
  (2019).
\newblock Transition from convection rolls to large-scale cellular structures
  in turbulent {R}ayleigh-{B}{\'e}nard convection in a liquid metal layer.
\newblock {\em Phys.~Rev.~Fluids}, 4(3):033501.

\bibitem[Anders et~al., 2019]{Anders19}
Anders, E., Manduca, C., Brown, B., Oishi, J., and Vasil, G. (2019).
\newblock Predicting the rossby number in convective experiments.
\newblock {\em Astrophys.~J.}, 872(2):138.

\bibitem[Aubert et~al., 2001]{Aubert01}
Aubert, J., Brito, D., Nataf, H.-C., Cardin, P., and Masson, J.-P. (2001).
\newblock A systematic experimental study of rapidly rotating spherical
  convection in water and liquid gallium.
\newblock {\em Phys.~Earth Planet.~Inter.}, 128(1-4):51--74.

\bibitem[Aurnou et~al., 2018]{Aurnou2018}
Aurnou, J., Bertin, V., Grannan, A., Horn, S., and Vogt, T. (2018).
\newblock Rotating thermal convection in liquid gallium: multi-modal flow,
  absent steady columns.
\newblock {\em J.~Fluid Mech.}, 846:846--876.

\bibitem[Aurnou et~al., 2007]{Aurnou07}
Aurnou, J., Heimpel, M., and Wicht, J. (2007).
\newblock The effects of vigorous mixing in a convective model of zonal flow on
  the {I}ce {G}iants.
\newblock {\em Icarus}, 190(1):110--126.

\bibitem[Barker et~al., 2014]{Barker14}
Barker, A., Dempsey, A., and Lithwick, Y. (2014).
\newblock Theory and simulations of rotating convection.
\newblock {\em Astrophys.~J.}, 791(1):13.

\bibitem[Bouillaut et~al., 2019]{Bouillaut19}
Bouillaut, V., Lepot, S., Auma{\^\i}tre, S., and Gallet, B. (2019).
\newblock Transition to the ultimate regime in a radiatively driven convection
  experiment.
\newblock {\em J.~Fluid Mech.}, 861.

\bibitem[Brown and Ahlers, 2006]{Brown06Coriolis}
Brown, E. and Ahlers, G. (2006).
\newblock {Effect of the Earth’s Coriolis force on the large-scale
  circulation of turbulent Rayleigh-B{\'e}nard convection}.
\newblock {\em Phys.~Fluids}, 18(12):125108.

\bibitem[Calkins, 2018]{Calkins18}
Calkins, M. (2018).
\newblock Quasi-geostrophic dynamo theory.
\newblock {\em Phys.~Earth Planet.~Inter.}, 276:182--189.

\bibitem[Calkins et~al., 2015a]{Calkins15rapids}
Calkins, M., Hale, K., Julien, K., Nieves, D., Driggs, D., and Marti, P.
  (2015a).
\newblock The asymptotic equivalence of fixed heat flux and fixed temperature
  thermal boundary conditions for rapidly rotating convection.
\newblock {\em J.~Fluid Mech.}, 784.

\bibitem[Calkins et~al., 2015b]{Calkins15}
Calkins, M., Julien, K., Tobias, S., and Aurnou, J. (2015b).
\newblock A multiscale dynamo model driven by quasi-geostrophic convection.
\newblock {\em J.~Fluid Mech.}, 780:143--166.

\bibitem[Calkins et~al., 2015c]{Calkins15Rapid}
Calkins, M.~A., Hale, K., Julien, K., Nieves, D., Driggs, D., and Marti, P.
  (2015c).
\newblock The asymptotic equivalence of fixed heat flux and fixed temperature
  thermal boundary conditions for rapidly rotating convection.
\newblock {\em J.~Fluid Mech.}, 784.

\bibitem[Cao et~al., 2018]{Cao18}
Cao, H., Yadav, R., and Aurnou, J. (2018).
\newblock Geomagnetic polar minima do not arise from steady meridional
  circulation.
\newblock {\em Proc.~Natl.~Acad.~Sci.~USA}, 115(44):11186--11191.

\bibitem[Chandrasekhar, 1961]{Chandra61}
Chandrasekhar, S. (1961).
\newblock {\em Hydrodynamic and Hydromagnetic Stability}.
\newblock Dover.

\bibitem[Chandrasekhar and Elbert, 1955]{ChandraElbert55}
Chandrasekhar, S. and Elbert, D. (1955).
\newblock The instability of a layer of fluid heated below and subject to
  coriolis forces. ii.
\newblock {\em Proc.~Roy.~Soc.~Lond.~A}, 231(1185):198--210.

\bibitem[Cheng and Aurnou, 2016]{Cheng16}
Cheng, J. and Aurnou, J. (2016).
\newblock Tests of diffusion-free scaling behaviors in numerical dynamo
  datasets.
\newblock {\em Earth Planet.~Sci.~Lett.}, 436:121--129.

\bibitem[Cheng et~al., 2018]{Cheng18}
Cheng, J., Aurnou, J., Julien, K., and Kunnen, R. (2018).
\newblock A heuristic framework for next-generation models of geostrophic
  convective turbulence.
\newblock {\em Geophys.~Astrophys.~Fluid Dyn.}, 112(4):277--300.

\bibitem[Cheng et~al., 2020]{Cheng20}
Cheng, J., Madonia, M., Aguirre~Guzmán, A., and Kunnen, R. (2020).
\newblock Laboratory exploration of heat transfer regimes in rapidly rotating
  turbulent convection.
\newblock {\em Phys.~Rev.~Lett.}, XXX:Submitted.

\bibitem[Cheng et~al., 2015]{Cheng15}
Cheng, J.~S., Stellmach, S., Ribeiro, A., Grannan, A., King, E.~M., and Aurnou,
  J.~M. (2015).
\newblock Laboratory-numerical models of rapidly rotating convection in
  planetary cores.
\newblock {\em Geophys.~J.~Int.}, 201:1--17.

\bibitem[Chill{\`a} and Schumacher, 2012]{Chilla2012}
Chill{\`a}, F. and Schumacher, J. (2012).
\newblock {New perspectives in turbulent Rayleigh-B{\'e}nard convection}.
\newblock {\em Eur.~Phys.~J.~E}, 35(7):1--25.

\bibitem[Christensen, 2002]{Christensen02}
Christensen, U. (2002).
\newblock Zonal flow driven by strongly supercritical convection in rotating
  spherical shells.
\newblock {\em J.~Fluid Mech.}, 470:115.

\bibitem[Christensen, 2010]{Christensen10}
Christensen, U. (2010).
\newblock Dynamo scaling laws and applications to the planets.
\newblock {\em Space Sci.~Rev.}, 152(1-4):565 -- 590.

\bibitem[Christensen and Aubert, 2006]{Christensen06}
Christensen, U. and Aubert, J. (2006).
\newblock Scaling properties of convection-driven dynamos in rotating spherical
  shells and application to planetary magnetic fields.
\newblock {\em Geophys.~J.~Int.}, 166(1):97--114.

\bibitem[Dawes, 2001]{Dawes01}
Dawes, J. (2001).
\newblock Rapidly rotating thermal convection at low {P}randtl number.
\newblock {\em J.~Fluid Mech.}, 428:61--80.

\bibitem[Deardorff, 1970]{Deardorff70}
Deardorff, J.~W. (1970).
\newblock Convective velocity and temperature scales for the unstable planetary
  boundary layer and for rayleigh convection.
\newblock {\em Journal of the atmospheric sciences}, 27(8):1211--1213.

\bibitem[Diamond et~al., 2009]{Diamond09}
Diamond, P., McDevitt, C., G{\"u}rcan, O., Hahm, T., Wang, W., Yoon, E., Holod,
  I., Lin, Z., Naulin, V., and Singh, R. (2009).
\newblock Physics of non-diffusive turbulent transport of momentum and the
  origins of spontaneous rotation in tokamaks.
\newblock {\em Nuclear Fusion}, 49(4):045002.

\bibitem[Doering, 2020]{Doering20}
Doering, C. (2020).
\newblock Turning up the heat in turbulent thermal convection.
\newblock {\em Proc.~Natl.~Acad.~Sci.~USA}, 117(18):9671--9673.

\bibitem[Doering et~al., 2019]{Doering19}
Doering, C., Toppaladoddi, S., and Wetlaufer, J. (2019).
\newblock Abscence of evidence for the ultimate regime in two-dimensional
  {R}ayleigh-{B}\'enard convection.
\newblock {\em Phys.~Rev.~Lett.}, 123(25):259401.

\bibitem[Ecke, 2015]{Ecke15}
Ecke, R. (2015).
\newblock Scaling of heat transport near onset in rapidly rotating convection.
\newblock {\em Phys.~Lett.~A}, 379(37):2221--2223.

\bibitem[Ecke and Niemela, 2014]{Ecke14}
Ecke, R. and Niemela, J. (2014).
\newblock Heat transport in the geostrophic regime of rotating
  {R}ayleigh-{B}{\'e}nard convection.
\newblock {\em Phys.~Rev.~Lett.}, 113(11):114301.

\bibitem[Featherstone and Hindman, 2016]{Featherstone16spectral}
Featherstone, N. and Hindman, B. (2016).
\newblock The spectral amplitude of stellar convection and its scaling in the
  high-{R}ayleigh-number regime.
\newblock {\em Astrophys.~J.}, 818(1):32.

\bibitem[Featherstone and Miesch, 2015]{Featherstone15}
Featherstone, N. and Miesch, M. (2015).
\newblock Meridional circulation in solar and stellar convection zones.
\newblock {\em Astrophys.~J.}, 804(1):67.

\bibitem[Gastine et~al., 2016]{Gastine16}
Gastine, T., Wicht, J., and Aubert, J. (2016).
\newblock Scaling regimes in spherical shell rotating convection.
\newblock {\em J.~Fluid Mech.}, 808:690 -- 732.

\bibitem[Gastine et~al., 2013a]{Gastine13}
Gastine, T., Wicht, J., and Aurnou, J. (2013a).
\newblock Zonal flow regimes in rotating anelastic spherical shells: an
  application to giant planets.
\newblock {\em Icarus}, 225(1):156--172.

\bibitem[Gastine et~al., 2015]{Gastine15}
Gastine, T., Wicht, J., and Aurnou, J. (2015).
\newblock Turbulent {R}ayleigh--b{\'e}nard convection in spherical shells.
\newblock {\em J.~Fluid Mech.}, 778:721--764.

\bibitem[Gastine et~al., 2013b]{gastine2013solar}
Gastine, T., Yadav, R., Morin, J., Reiners, A., and Wicht, J. (2013b).
\newblock From solar-like to antisolar differential rotation in cool stars.
\newblock {\em Month.~Not.~Roy.~Astron.~Soc.}, 438(1):L76--L80.

\bibitem[Gastine et~al., 2013c]{Gastine13MNRAS}
Gastine, T., Yadav, R.~K., Morin, J., Reiners, A., and Wicht, J. (2013c).
\newblock {From solar-like to antisolar differential rotation in cool stars}.
\newblock {\em Month.~Not.~Roy.~Astron.~Soc.}, 438(1):L76--L80.

\bibitem[Gillet and Jones, 2006]{Gillet06}
Gillet, N. and Jones, C. (2006).
\newblock The quasi-geostrophic model for rapidly rotating spherical convection
  outside the tangent cylinder.
\newblock {\em J.~Fluid Mech.}, 554:343.

\bibitem[Gilman, 1977]{Gilman77}
Gilman, P. (1977).
\newblock Nonlinear dynamics of {B}oussinesq convection in a deep rotating
  spherical shell-i.
\newblock {\em Geophys.~Astrophys.~Fluid Dyn.}, 8(1):93--135.

\bibitem[Glatzmaier, 2002]{GAG02}
Glatzmaier, G. (2002).
\newblock Geodynamo simulations - how realistic are they?
\newblock {\em Annu.~Rev.~Earth Planet.~Sci.}, 30:237 -- 257.

\bibitem[Glazier et~al., 1999]{Glazier99}
Glazier, J., Segawa, T., Naert, A., and Sano, M. (1999).
\newblock Evidence against ‘ultrahard’ thermal turbulence at very high
  {R}ayleigh numbers.
\newblock {\em Nature}, 398(6725):307.

\bibitem[Greenspan, 1968]{Greenspan68}
Greenspan, H. (1968).
\newblock {\em The theory of rotating fluids}.
\newblock CUP Archive.

\bibitem[Grooms and Whitehead, 2014]{Grooms14}
Grooms, I. and Whitehead, J. (2014).
\newblock Bounds on heat transport in rapidly rotating {R}ayleigh--{B}{\'e}nard
  convection.
\newblock {\em Nonlinearity}, 28(1):29.

\bibitem[Grossmann and Lohse, 2000]{Grossmann00}
Grossmann, S. and Lohse, D. (2000).
\newblock {Scaling in thermal convection: A unifying theory}.
\newblock {\em J.\ Fluid Mech.}, 407:27--56.

\bibitem[Grossmann and Lohse, 2004]{Grossmann04}
Grossmann, S. and Lohse, D. (2004).
\newblock {Fluctuations in turbulent {R}ayleigh-{B}{\'e}nard convection: the
  role of plumes}.
\newblock {\em Phys. Fluids}, 16:4462.

\bibitem[Grossmann and Lohse, 2011]{Grossmann2011}
Grossmann, S. and Lohse, D. (2011).
\newblock Multiple scaling in the ultimate regime of thermal convection.
\newblock {\em Phys. Fluids}, 23:045108.

\bibitem[Guervilly and Cardin, 2016]{Guervilly16}
Guervilly, C. and Cardin, P. (2016).
\newblock Subcritical convection of liquid metals in a rotating sphere using a
  quasi-geostrophic model.
\newblock {\em J.~Fluid Mech.}, 808:61--89.

\bibitem[Guervilly et~al., 2019]{Guervilly19}
Guervilly, C., Cardin, P., and Schaeffer, N. (2019).
\newblock Turbulent convective length scale in planetary cores.
\newblock {\em Nature}, 570(7761):368.

\bibitem[G{\"u}rcan and Diamond, 2015]{Gurcan15}
G{\"u}rcan, {\"O}. and Diamond, P. (2015).
\newblock Zonal flows and pattern formation.
\newblock {\em Journal Phys.~A: Math. Theor.}, 48(29):293001.

\bibitem[Hart and Ohlsen, 1999]{Hart99}
Hart, J. and Ohlsen, D. (1999).
\newblock On the thermal offset in turbulent rotating convection.
\newblock {\em Phys.~Fluids}, 11(8):2101--2107.

\bibitem[Hawkins et~al., 2020]{Hawkins20}
Hawkins, E., Cheng, J., Pilegard, T., Stellmach, S., and Aurnou, J. (2020).
\newblock Separate regime transitions for heat transfer and bulk dynamics in
  models of planetary core convection.
\newblock {\em Geophys.~J.~Int.}, XXX:Submitted.

\bibitem[He et~al., 2012]{He12}
He, X., Funfschilling, D., Nobach, H., Bodenschatz, E., and Ahlers, G. (2012).
\newblock Transition to the ultimate state of turbulent {R}ayleigh-{B}{\'e}nard
  convection.
\newblock {\em Phys.~Rev.~Lett.}, 108(2):024502.

\bibitem[Horn and Aurnou, 2018]{Horn18}
Horn, S. and Aurnou, J. (2018).
\newblock Regimes of {C}oriolis-centrifugal convection.
\newblock {\em Phys.~Rev.~Lett.}, 120(20):204502.

\bibitem[Horn and Aurnou, 2019]{Horn19}
Horn, S. and Aurnou, J. (2019).
\newblock "rotating convection with centrifugal buoyancy: Numerical predictions
  for laboratory experiments".
\newblock {\em PRF}, 4(7):073501.

\bibitem[Horn and Schmid, 2017]{Horn17}
Horn, S. and Schmid, P. (2017).
\newblock Prograde, retrograde, and oscillatory modes in rotating
  {R}ayleigh-{B}{\'e}nard convection.
\newblock {\em J.~Fluid Mech.}, 831:182--211.

\bibitem[Horn and Shishkina, 2014]{Horn14}
Horn, S. and Shishkina, O. (2014).
\newblock "rotating non-oberbeck--boussinesq rayleigh--b{\'e}nard convection in
  water".
\newblock {\em Phys.~Fluids}, 26(5):055111.

\bibitem[Horn and Shishkina, 2015]{Horn15}
Horn, S. and Shishkina, O. (2015).
\newblock Toroidal and poloidal energy in rotating {R}ayleigh--{B}{\'e}nard
  convection.
\newblock {\em J.~Fluid Mech.}, 762:232--255.

\bibitem[Ingersoll and Pollard, 1982]{Ingersoll82}
Ingersoll, A. and Pollard, D. (1982).
\newblock Motion in the interiors and atmospheres of {J}upiter and {S}aturn:
  Scale analysis, anelastic equations, barotropic stability criterion.
\newblock {\em Icarus}, 52(1):62--80.

\bibitem[Jones et~al., 2000]{Jones00}
Jones, C., Soward, A., and Mussa, A. (2000).
\newblock The onset of thermal convection in a rapidly rotating sphere.
\newblock {\em J.~Fluid Mech.}, 405:157–179.

\bibitem[Jones, 2011]{Jones11}
Jones, C.~A. (2011).
\newblock Planetary magnetic fields and fluid dynamos.
\newblock {\em Ann.~Rev.~Fluid Mech.}, 43:583--614.

\bibitem[Julien et~al., 2016]{Julien16}
Julien, K., Aurnou, J., Calkins, M., Knobloch, E., Marti, P., Stellmach, S.,
  and Vasil, G. (2016).
\newblock A nonlinear model for rotationally constrained convection with
  {E}kman pumping.
\newblock {\em J.~Fluid Mech.}, 798:50--87.

\bibitem[Julien and Knobloch, 1997]{Julien97}
Julien, K. and Knobloch, E. (1997).
\newblock Fully nonlinear oscillatory convection in a rotating layer.
\newblock {\em Phys.~Fluids}, 9(7):1906--1913.

\bibitem[Julien and Knobloch, 1998]{Julien98}
Julien, K. and Knobloch, E. (1998).
\newblock {Strongly nonlinear convection cells in a rapidly rotating fluid
  layer: the tilted $f$-plane}.
\newblock {\em J.~Fluid Mech.}, 360:141--178.

\bibitem[Julien and Knobloch, 1999]{Julien99}
Julien, K. and Knobloch, E. (1999).
\newblock Fully nonlinear three-dimensional convection in a rapidly rotating
  layer.
\newblock {\em Phys.~Fluids}, 11(6):1469--1483.

\bibitem[Julien et~al., 2012a]{Julien12}
Julien, K., Knobloch, E., Rubio, A., and Vasil, G. (2012a).
\newblock Heat transport in low-{R}ossby-number {R}ayleigh-{B}\'enard
  convection.
\newblock {\em Phys.~Rev.~Lett.}, 109(25):254503.

\bibitem[Julien et~al., 1996]{Julien96}
Julien, K., Legg, S., McWilliams, J., and Werne, J. (1996).
\newblock Rapidly rotating turbulent {R}ayleigh-{B}{\'e}nard convection.
\newblock {\em J.~Fluid Mech.}, 322:243--273.

\bibitem[Julien et~al., 2012b]{Julien12gafd}
Julien, K., Rubio, A., Grooms, I., and Knobloch, E. (2012b).
\newblock Statistical and physical balances in low {R}ossby number
  {R}ayleigh--{B}{\'e}nard convection.
\newblock {\em Geophys.~Astrophys.~Fluid Dyn.}, 106(4-5):392--428.

\bibitem[Kaplan et~al., 2017]{Kaplan17}
Kaplan, E., Schaeffer, N., Vidal, J., and Cardin, P. (2017).
\newblock Subcritical thermal convection of liquid metals in a rapidly rotating
  sphere.
\newblock {\em Phys.~Rev.~Lett.}, 119(9):094501.

\bibitem[King and Aurnou, 2013]{KingAurnou13}
King, E. and Aurnou, J. (2013).
\newblock Turbulent convection in liquid metal with and without rotation.
\newblock {\em Proc.~Natl.~Acad.~Sci.~USA}, 110(17):6688--6693.

\bibitem[King and Aurnou, 2015]{King15}
King, E. and Aurnou, J. (2015).
\newblock Magnetostrophic balance as the optimal state for turbulent
  magnetoconvection.
\newblock {\em Proc.~Natl.~Acad.~Sci.~USA}, 112(4):990--994.

\bibitem[King et~al., 2009]{King09}
King, E., Stellmach, S., Noir, J., Hansen, U., and Aurnou, J. (2009).
\newblock Boundary layer control of rotating convection systems.
\newblock {\em Nature}, 457(7227):301--304.

\bibitem[King and Buffett, 2013]{KingBuffett13}
King, E.~M. and Buffett, B.~A. (2013).
\newblock Flow speeds and length scales in geodynamo models: The role of
  viscosity.
\newblock {\em Earth Planet.~Sci.~Lett.}, 371--372:156--162.

\bibitem[Kunnen et~al., 2008]{Kunnen08}
Kunnen, R., Clercx, H., and Geurts, B. (2008).
\newblock Breakdown of large-scale circulation in turbulent rotating
  convection.
\newblock {\em Europhys. Lett.}, 84(2):24001.

\bibitem[Lepot et~al., 2018]{Lepot18}
Lepot, S., Auma{\^\i}tre, S., and Gallet, B. (2018).
\newblock Radiative heating achieves the ultimate regime of thermal convection.
\newblock {\em Proc.~Natl.~Acad.~Sci.~USA}, 115(36):8937--8941.

\bibitem[Liu and Ecke, 1997]{Ecke97}
Liu, Y. and Ecke, R. (1997).
\newblock {Heat transport scaling in turbulent Rayleigh-B{\'e}nard convection:
  effects of rotation and Prandtl number}.
\newblock {\em Phys.~Rev.~Lett.}, 79(12):2257.

\bibitem[Lohse and Toschi, 2003]{Lohse03}
Lohse, D. and Toschi, F. (2003).
\newblock Ultimate state of thermal convection.
\newblock {\em Phys.~Rev.~Lett.}, 90(3):034502.

\bibitem[Mabuchi et~al., 2015]{Mabuchi15}
Mabuchi, J., Masada, Y., and Kageyama, A. (2015).
\newblock Differential rotation in magnetized and non-magnetized stars.
\newblock {\em Astrophys.~J.}, 806(1):10.

\bibitem[Maffei et~al., 2020]{Maffei20}
Maffei, S., Julien, K., Marti, P., and Calkins, M. (2020).
\newblock Systematic investigation of the inverse cascade and flow speeds in
  rapidly rotating {R}ayleigh-{B}{\'e}nard convection.
\newblock {\em J.~Fluid Mech.}, XXX:Submitted.

\bibitem[Marti et~al., 2016]{Marti16}
Marti, P., Calkins, M., and Julien, K. (2016).
\newblock A computationally efficient spectral method for modeling core
  dynamics.
\newblock {\em Geochemistry, Geophysics, Geosystems}, 17(8):3031--3053.

\bibitem[Maxworthy and Narimousa, 1994]{Maxworthy94}
Maxworthy, T. and Narimousa, S. (1994).
\newblock Unsteady, turbulent convection into a homogeneous, rotating fluid,
  with oceanographic applications.
\newblock {\em J.~Phys.~Ocean.}, 24(5):865--887.

\bibitem[McWilliams, 2006]{McWilliams06}
McWilliams, J. (2006).
\newblock {\em Fundamentals of {G}eophysical {F}luid {D}ynamics}.
\newblock [Cambridge], Cambridge University Press.

\bibitem[Pandey et~al., 2018]{Pandey18}
Pandey, A., Scheel, J., and Schumacher, J. (2018).
\newblock Turbulent superstructures in {R}ayleigh-{B}{\'e}nard convection.
\newblock {\em Nature Comm.}, 9(1):2118.

\bibitem[Plumley and Julien, 2019]{Plumley19}
Plumley, M. and Julien, K. (2019).
\newblock Scaling laws in {R}ayleigh-{B}\'enard convection.
\newblock {\em Earth and Space Science}, 34:1--13.

\bibitem[Plumley et~al., 2016]{Plumley16}
Plumley, M., Julien, K., Marti, P., and Stellmach, S. (2016).
\newblock The effects of {E}kman pumping on quasi-geostrophic
  {R}ayleigh--{B}{\'e}nard convection.
\newblock {\em J.~Fluid Mech.}, 803:51--71.

\bibitem[Plumley et~al., 2017]{Plumley17}
Plumley, M., Julien, K., Marti, P., and Stellmach, S. (2017).
\newblock Sensitivity of rapidly rotating {R}ayleigh-{B}{\'e}nard convection to
  {E}kman pumping.
\newblock {\em Phys.~Rev.~Fluids}, 2(9):094801.

\bibitem[Rieutord and Rincon, 2010]{Rieutord2010}
Rieutord, M. and Rincon, F. (2010).
\newblock The {S}un’s supergranulation.
\newblock {\em Living Reviews in Solar Physics}, 7(1):2.

\bibitem[Roberts and King, 2013]{RobertsKing13}
Roberts, P.~H. and King, E.~M. (2013).
\newblock On the genesis of the {E}arth's magnetism.
\newblock {\em Rev.~Prog.~Phys.}, 76.

\bibitem[Rossby, 1969]{Rossby69}
Rossby, H. (1969).
\newblock {A study of B{\'e}nard convection with and without rotation}.
\newblock {\em J.~Fluid Mech.}, 36(2):309--335.

\bibitem[Sakievich et~al., 2016]{Adrian16}
Sakievich, P., Peet, Y., and Adrian, R. (2016).
\newblock Large-scale thermal motions of turbulent rayleigh--b{\'e}nard
  convection in a wide aspect-ratio cylindrical domain.
\newblock {\em International Journal of Heat and Fluid Flow}, 61:183--196.

\bibitem[Soderlund et~al., 2012]{Soderlund12}
Soderlund, K., King, E., and Aurnou, J. (2012).
\newblock The influence of magnetic fields in planetary dynamo models.
\newblock {\em Earth Planet.~Sci.~Lett.}, 333:9--20.

\bibitem[Soderlund et~al., 2014]{Soderlund14}
Soderlund, K., Schmidt, B., Wicht, J., and Blankenship, D. (2014).
\newblock Ocean-driven heating of {E}uropa’s icy shell at low latitudes.
\newblock {\em Nature Geosci.}, 7(1):16--19.

\bibitem[Sparrow et~al., 1964]{Sparrow64}
Sparrow, E., Goldstein, R., and Jonsson, V. (1964).
\newblock Thermal instability in a horizontal fluid layer: Effect of boundary
  conditions and non-linear temperature profile.
\newblock {\em J.~Fluid Mech.}, 18(4):513--528.

\bibitem[Spiegel, 1963]{Spiegel63}
Spiegel, E. (1963).
\newblock A generalization of the mixing-length theory of turbulent convection.
\newblock {\em Astrophys.~J.}, 138:216.

\bibitem[Spiegel, 1971]{Spiegel71}
Spiegel, E.~A. (1971).
\newblock Convection in stars i. basic boussinesq convection.
\newblock {\em Ann.~Rev.~Astron.~Astrophys.}, 9(1):323--352.

\bibitem[Sprague et~al., 2006]{Sprague06}
Sprague, M., Julien, K., Knobloch, E., and Werne, J. (2006).
\newblock Numerical simulation of an asymptotically reduced system for
  rotationally constrained convection.
\newblock {\em J.~Fluid Mech.}, 551:141--174.

\bibitem[Stellmach and Hansen, 2004]{Stellmach04}
Stellmach, S. and Hansen, U. (2004).
\newblock Cartesian convection driven dynamos at low {E}kman number.
\newblock {\em Phys.~Rev.~E}, 70(5):056312.

\bibitem[Stevens et~al., 2013]{Stevens13}
Stevens, R. J. A.~M., van~der Poel, E.~P., Grossmann, S., and Lohse, D. (2013).
\newblock The unifying theory of scaling in thermal convection: the updated
  prefactors.
\newblock {\em J.~Fluid Mech.}, 730:295--308.

\bibitem[Stevenson, 1979]{Stevenson79}
Stevenson, D. (1979).
\newblock Turbulent thermal convection in the presence of rotation and a
  magnetic field: {A} heuristic theory.
\newblock {\em Geophys.~Astrophys.~Fluid Dyn.}, 12:139--169.

\bibitem[Verma, 2018]{Verma18}
Verma, M. (2018).
\newblock {\em "Physics of Buoyant Flows: From Instabilities to Turbulence"}.
\newblock World Scientific.

\bibitem[Vogt et~al., 2020]{Vogt20}
Vogt, T., Horn, S., and Aurnou, J. (2020).
\newblock Thermal-inertial oscillatory flows in liquid metal rotating
  convection.
\newblock {\em J.~Fluid Mech.}, XX(XX):XX.

\bibitem[Vogt et~al., 2018]{Vogt18}
Vogt, T., Horn, S., Grannan, A., and Aurnou, J. (2018).
\newblock Jump rope vortex in liquid metal convection.
\newblock {\em Proc.~Natl.~Acad.~Sci.~USA}, 115(50):12674--12679.

\bibitem[Von~Hardenberg et~al., 2008]{vonHardenberg08}
Von~Hardenberg, J., Parodi, A., Passoni, G., Provenzale, A., and Spiegel, E.
  (2008).
\newblock Large-scale patterns in rayleigh--b{\'e}nard convection.
\newblock {\em Physics Letters A}, 372(13):2223--2229.

\bibitem[Weiss and Ahlers, 2011]{Weiss11}
Weiss, S. and Ahlers, G. (2011).
\newblock Heat transport by turbulent rotating {R}ayleigh--{B}{\'e}nard
  convection and its dependence on the aspect ratio.
\newblock {\em J.~Fluid Mech.}, 684:407--426.

\bibitem[Wood et~al., 2013]{Wood13}
Wood, T., Garaud, P., and Stellmach, S. (2013).
\newblock A new model for mixing by double-diffusive convection
  (semi-convection). ii. {T}he transport of heat and composition through
  layers.
\newblock {\em Astrophys.~J.}, 768(2):157.

\bibitem[Zhang and Liao, 2017]{Zhang17}
Zhang, K. and Liao, X. (2017).
\newblock {\em Theory and Modeling of Rotating Fluids: Convection, Inertial
  Waves and Precession}.
\newblock Cambridge University Press.

\bibitem[Zhong et~al., 2017]{JQZ17}
Zhong, J., Li, H., and Wang, X. (2017).
\newblock Enhanced azimuthal rotation of the large-scale flow through
  stochastic cessations in turbulent rotating convection with large {R}ossby
  numbers.
\newblock {\em Phys.~Rev.~Fluids}, 2(4):044602.

\bibitem[Zhong et~al., 2009]{Zhong09}
Zhong, J., Stevens, R., Clercx, H., Verzicco, R., Lohse, D., and Ahlers, G.
  (2009).
\newblock Prandtl-, {R}ayleigh-, and {R}ossby-number dependence of heat
  transport in turbulent rotating {R}ayleigh-{B}{\'e}nard convection.
\newblock {\em Phys.~Rev.~Lett.}, 102(4):044502.

\bibitem[Zhu et~al., 2019]{Zhu19}
Zhu, X., Mathai, V., Stevens, R., Verzicco, R., and Lohse, D. (2019).
\newblock Reply to ``{A}bscence of evidence for the ultimate regime in
  two-dimensional {R}ayleigh-{B}\'enard convection''.
\newblock {\em Phys.~Rev.~Lett.}, 123(25):259402.

\bibitem[Z{\"u}rner et~al., 2019]{Zurner19}
Z{\"u}rner, T., Schindler, F., Vogt, T., Eckert, S., and Schumacher, J. (2019).
\newblock Combined measurement of velocity and temperature in liquid metal
  convection.
\newblock {\em J.~Fluid Mech.}, 876:1108--1128.

\end{thebibliography}
\end{document}